\documentclass[AMA,STIX1COL]{WileyNJD-v2}

\usepackage[colorinlistoftodos]{todonotes}

\usepackage{siunitx}
\usepackage{chngcntr}

\usepackage{centernot}
\usepackage{xcolor}
\usepackage{amssymb}
\usepackage{pifont}
\usepackage{multirow}
\usepackage{lscape}
\usepackage{longtable}
\usepackage{amsmath}
\usepackage{arydshln}

\usepackage{subcaption}

\articletype{MAIN PAPER}%
\received{<day> <Month>, <year>}
\revised{<day> <Month>, <year>}
\accepted{<day> <Month>, <year>}

\usepackage{tikz}
\usetikzlibrary{shapes.geometric, arrows}

\tikzstyle{question} = [rectangle, rounded corners, minimum width=3cm, minimum height=1cm, text centered, draw=black, fill=red!30]
\tikzstyle{process} = [rectangle, minimum width=3cm, minimum height=1cm, text centered, draw=black, fill=none]
\tikzstyle{arrow} = [thick,->,>=stealth]

\begin{document}

\title{Overview and practical recommendations on using Shapley Values for identifying predictive biomarkers via CATE modeling}

\author[1]{David Svensson*}

\author[1]{Erik Hermansson}

\author[2]{Nikolaos Nikolaou}

\author[3]{Konstantinos Sechidis}

\author[4]{Ilya Lipkovich}

\authormark{AUTHOR ONE \textsc{et al}}

\address[1]{ \orgname{AstraZeneca}, \orgaddress{\city{Gothenburg},  \country{Sweden}}}
\address[2]{ \orgname{UCL}, \orgaddress{\city{London}, \country{UK}}}
\address[3]{ \orgname{Novartis}, \orgaddress{\city{Basel}, \state{Basel}, \country{Switzerland}}}
\address[4]{ \orgname{Eli Lilly and Company}, \orgaddress{\city{Indianapolis}, \state{Indiana}, \country{USA}}}

\corres{*David Svensson, \email{david.j.svensson@gmail.com}}

\presentaddress{Present address}

\abstract[Abstract]{
In recent years, two parallel research trends have emerged in machine learning, yet their intersections remain largely unexplored. On one hand, there has been a significant increase in literature focused on Individual Treatment Effect (ITE) modeling, particularly targeting the Conditional Average Treatment Effect (CATE) using meta-learner techniques. These approaches often aim to identify causal effects from observational data. On the other hand, the field of Explainable Machine Learning (XML) has gained traction, with various approaches developed to explain complex models and make their predictions more interpretable. A prominent technique in this area is Shapley Additive Explanations (SHAP), which has become mainstream in data science for analyzing supervised learning models. However, there has been limited exploration of SHAP’s application in identifying predictive biomarkers through CATE models, a crucial aspect in pharmaceutical precision medicine. We address inherent challenges associated with the SHAP concept in multi-stage CATE strategies and introduce a surrogate estimation approach that is agnostic to the choice of CATE strategy, effectively reducing computational burdens in high-dimensional data. Using this approach, we conduct simulation benchmarking to evaluate the ability to accurately identify biomarkers using SHAP values derived from various CATE meta-learners and Causal Forest.}

\keywords{Treatment Effect Heterogeneity; Precision Medicine;  Individual Treatment Effects; Causal Inference; Machine Learning; Conditional Average Treatment Effect; Prognostic and Predictive biomarkers; Shapley Values; SHAP; XML; Discovery Rates; Simulation }

\maketitle


\section{Introduction }\label{sec1}
In recent years, the development of methodologies for estimating causal effects has attracted significant attention. Applications of these methods span diverse scientific fields, including economics, social sciences, drug development, and precision medicine.\cite{JacobsCATE, UPLIFT2016, Kennedy2020, Knaus, ZHANG} The objectives of these studies are wide-ranging. Examples include identifying potential customers for a new product, evaluating the effects of new policies in society (e.g., determining who is more likely to comply or estimating the outcomes of a policy on different groups), and assessing the impacts of novel medical treatments (e.g., for better tailoring treatment regimes to patients or for identifying which patients should be prioritized when resources are limited). While specifics may differ across various fields, there is a significant interest in utilizing causal inference techniques with machine learning (ML), to assess causal effects. In this paper, we focus on causal effects of binary interventions, commonly referred to as ``treatments''. 

One popular causal estimand in this setting is the Conditional Average Treatment Effect (CATE). CATE captures the difference in expected outcomes if the same patient were to hypothetically be treated with two different treatments in ``parallel worlds'' as a function of covariates.\cite{ILDS} The term ``conditional''  refers to our interest in evaluating the treatment effect averaged over the experimental units (i.e., subjects in our case) having the same values on a subset of their covariates. The granularity may vary from conditioning on subpopulations defined by a small number of demographic variables to conditioning on high-dimensional biomarkers (to use in a clinical setting). Modern approaches to this problem differ substantially in their overall modeling strategies as well as their implementation details. One popular group of these, referred to as meta-learners, \cite{Chen2017,Kuenzel2019,Jacob2020,Huling2021} combines ML models targeting potential outcomes within a causal inference framework. Even though the general strategy is the same, existing methods can differ significantly in how they construct multi-stage strategies involving numerous modeling steps.\cite{Knaus, Kennedy2020} The various meta-learner implementations diverge on aspects such as the choice of base learners, and the specifics of model fitting (such as data splitting, relative weighting and regularization). Additionally, there are other methodologies that aim to directly estimate CATE via modifications of existing supervised learning techniques. Examples of these include Causal Forest\cite{AtheyWager2019}, Causal Boosting \cite{Powers2018}, and Bayesian Causal Forest \cite{HahnBARTCRAN}. 


In the context of drug development, evaluating potential treatment effect heterogeneity encompasses more than simply estimating individual treatment effects.\cite{Huang2021, TRALO, LOH13} It is also pivotal to identifying which biomarkers, if any, interact with the treatment, thus contributing to CATE.\cite{NeurIPS} As this knowledge forms the foundation of any precision medicine strategy, it necessitates insights into the likely biomarkers (and their functional relationships) associated with differential treatment effects. The importance of identifying predictive covariates has been highlighted in the recent framework, WATCH (Workflow to Assess Treatment Effect Heterogeneity), which is designed to address the challenges of investigating treatment effect heterogeneity in randomized clinical trials.\cite{sechidis2024} As highlighted by Sechidis et al.\cite{INFOPLUS}, basing treatment assignments on a biomarker that is merely prognostic (i.e. predicting outcomes under either treatment but ``canceling out'' in the conditional treatment effects) is likely to overestimate the benefits of the treatment for a subsection of the population, leading to undesirable consequences for various stakeholders (patients, sponsors and society in general).


There appears to be no unified approach for ranking baseline covariates based on their predictive likelihood of treatment effects. This contrasts with standard supervised machine learning\cite{ESL}, where numerous model-specific measures of variable importance exist. Recently, a new method has emerged in the data science field: Shapley Additive Explanations (SHAP).\cite{Lundberg2017} This approach applies the game-theoretical concept of the Shapley Value, originally designed to fairly allocate rewards among individuals after a collaborative effort, to machine learning models. In this context, the ``effort'' refers to the model's prediction, while the ``rewards'' relate to the contributions of the covariates to it. One of the key advantages of SHAP is its higher resolution compared to traditional summary measures of variable importance, such as the Gini index. This allows for deeper insights into trends and decision boundaries. Additionally, SHAPs possess desirable theoretical properties, which we will discuss further.

Despite the growing popularity of SHAP values in supervised machine learning, there has been limited attention to their application in causal inference and particularly in CATE modeling. To the best of our knowledge, no methodological paper has thoroughly explored this concept. The idea is straightforward and is articulated by Chernozhukov et al\cite[page 436]{ChernoBook}: \emph{``an alternative way to visualize the importance of the different covariates in changing the output of the CATE ensemble is by using the SHAP values''}. This is followed by a worked example where SHAP values are derived for a Q-aggregation-based stacked ensemble model for CATE, although details on the derivation are sparse. In the recent review paper by Mosca et al. \cite{mosca-etal-2022-shap}, a comprehensive overview of SHAP-related literature is provided, including a summary of how various methodological advancements interrelate (see their Table 1). It appears that only one of these papers, focusing on causal Shapley values, \cite{CausalSHAP} specifically addresses causal inference models, yet it does not explicitly cover CATE modeling. Notably, it acknowledges the potential computational burden involved, stating that it is \emph{``... considerably more expensive than computing marginal Shapley values.''} Furthermore, Hines et al. \cite{hines2022variable} introduced a measure of heterogeneous treatment effect variable importance called TE-VIM, which quantifies the change in the mean squared error when specific covariates are excluded from the model estimating CATE. They developed inference methods for TE-VIM using machine learning techniques, particularly doubly robust estimators of CATE. While they recognized the potential of SHAP values for this analysis, they chose not to pursue it further due to computational challenges. 
A likely reason for the limited focus on SHAP values in CATE modeling is that this task often involves multiple modeling steps, raising questions about how to effectively apply the SHAP concept. Additionally, model-agnostic (kernel-based) SHAP implementations struggle to handle datasets with a large number of covariates, leading to computational challenges that will be illustrated in Section~\ref{sec.compBurden}.

In this paper, we propose an alternative approach for deriving SHAP values of CATE estimators that addresses existing limitations while aligning with various subgroup detection strategies. We incorporate a surrogate modeling step, which regresses the estimated CATE against baseline covariates using a boosting model, from which SHAP importance is derived for each covariate. This two-step approach provides a novel and unified method for evaluating different CATE models based on their ability to accurately identify predictive covariates through SHAP rankings. This two-step approach has been utilized in other contexts. For example, it was used in the Virtual Twins method for subgroup selection\cite{Foster2011}, where a Classification and Regression Trees (CART) model is built on top of the estimated CATE to identify subpopulations with enhanced treatment effects. As noted by Man et al.\cite{BOOKBIOMARKER}, any supervised learning method beyond CART can be utilized for this secondary step. For example, ensemble techniques can generate a ranking of covariates, as discussed in the benchmarking paper by Huang et al.\cite{Huang2021}. The second step can also be repurposed to identify covariates with predictive power, i.e., those associated with estimated individual treatment effects from the first stage. For example, Man et al.\cite[see][Section 4.2]{BOOKBIOMARKER} suggest methods such as Random Forest\cite{Breiman1996}, BART\cite{BART}, and GUIDE\cite{loh2011classification} can be used in the second step. 


The goals of this tutorial paper are three-fold. First, we bring together diverse literature on evaluating heterogeneous treatment effects (HTE) via ML and explainable machine learning, specifically SHAP values, providing the reader with an up-to-date overview on theoretical developments and existing tools for both areas, as well as with our recommendations on the best strategies when using SHAP values to explain models aimed at estimating causal quantities (i.e., CATE) rather than traditional predictive models in supervised learning. Second, we include extensive simulation studies and investigate the operating characteristics of several popular CATE modeling choices (including T-, S-, X-, R- and DR-learning and Causal Forest) using several new performance metrics allowing an interested reader to further their knowledge on these methods and perhaps replicate some of our simulation designs in their own area using parameters tailored to the scenarios/data at hand. Third, we provide working examples of R code that can be readily used with minimal modifications by interested readers.    

The remainder of the paper is organized as follows: Section~\ref{sec.METH} introduces the methodology, beginning with a discussion of estimating CATE using meta-learners in Section~\ref{sec.CATE}, followed by an exploration of methods for explaining machine learning models in Section~\ref{sec.XML}, and ultimately combining these two areas to provide techniques for explaining CATE models in Section~\ref{sec.SHAP_CATE}. Section~\ref{sec.examples} presents example simulations that not only prepare the way for the larger benchmarking study but also serve to motivate the main insights of this work through simplified scenarios. These examples highlight important factors to consider in evaluations. A comprehensive simulation-based benchmarking is detailed in Section~\ref{sec.LSB}, accompanied by two case studies in Section~\ref{sec.CASE}. Finally, Section~\ref{sec.DISC} concludes the paper with a discussion of the results.   

\section{Methodology}\label{sec.METH}
This section is divided into three parts. Section~\ref{sec.CATE} focuses on estimating CATE using meta-learners. Section~\ref{sec.XML} delves into explaining machine learning models through SHAP values. Finally, Section~\ref{sec.SHAP_CATE} integrates these two areas, discussing techniques for explaining CATE models using SHAP values.


\subsection{Heterogeneous Treatment Effect Estimation\label{sec.CATE}}
In this section, we outline a framework for estimating and assessing potential heterogeneity in treatment effects, building upon the concept of \emph{potential outcomes}.\cite{Rubin1974} Research on estimating individual treatment effects is an active area of investigation, with numerous machine learning methods recently proposed for estimating Conditional Average Treatment Effects (CATE).\cite{JacobsCATE, Kuenzel2019, Knaus, Chernozhukov2020} We begin by introducing the necessary notation, a summary of which can be found in Table \ref{tab:symbols}. Let $( i = 1, \ldots, n)$ denote the index for each observation (i.e.  patient). Each patient is assumed to be exposed to one of two available treatments, represented by $(A_i \in {0, 1})$, where $0$ indicates the control arm and $1$ indicates the active treatment arm. While the general framework is applicable to observational data under standard assumptions, we will concentrate on randomized controlled trials (RCTs) in which patients are randomly assigned to one of the available treatments. This parallel design is commonly employed in Phase III drug development and is the focus of numerous studies.\cite{Lipkovich2017, Loh2019, Foster2011} Thus, $A_i$ follows a Bernoulli distribution with parameter $p_{rct}=Pr(A_i=1)$, where $A_i$  is independently distributed and $0 < p_{rct} < 1$. 

The potential outcomes are denoted $Y_i^{(a)}$ where $a=0, 1$ denotes treatment, i.e., it is assumed that $Y_i^{(0)}$ and $Y_i^{(1)}$ are the hypothetical outcomes if patient $i$ is exposed to treatment $A=0$ or $A=1$, respectively. For parallel trial designs only one potential outcome is observed for each patient and the other outcome  is counterfactual. Therefore, by consistency assumption, the observed outcome can be expressed via potential outcomes as    
\[Y_i= Y_i^{(0)}(1-A_i)+Y_i^{(1)}A_i.\] The causal effects  ITE $=Y_i^{(1)}-Y_i^{(0)}$ are hence fundamentally unobservable and must be inferred. 
In practice, estimating ITEs can be attempted by modeling them as functions of baseline covariates. Each patient is associated with a $p$-dimensional vector of covariates $\textbf{x}_i =(x_{i,1}, ..., x_{i,p})$ measured at baseline (i.e., prior to receiving any treatment). We often refer to this as biomarker data, as is common terminology for clinical  trial settings, or as features, as is common terminology in machine learning. For the remainder of this paper, the term \emph{covariates} will be used interchangeably with \emph{biomarkers}, \emph{features}, or \emph{input/independent variables}. 
 
A popular estimand of interest is the Conditional Average Treatment Effect (CATE) defined as  
\[\tau(\textbf{x}_i) = \mathbb{E}[Y^{(1)}\!-\!Y^{(0)}|X=\textbf{x}_i].\] 
The importance of CATE has been emphasized in the literature, for example, CATE is ``\emph{... the  best estimator of the ITE}'' \cite{Kuenzel2019} and ``\emph{... the primary target of almost all modern machine learning-based causal inference methods}'' \cite{IF}. From now, we will often suppress the index $i$ for readability. Furthermore, we denote
\[m_a(\textbf{x}) = \mathbb{E}[Y^{(a)}|X=\textbf{x}],\]
the mean outcome $Y$ for individuals receiving treatment $a$, conditioned on their characteristics $ \textbf{x}$ and hence $\tau(\textbf{x}) =  m_1(\textbf{x})-m_0(\textbf{x})$, a difference of two conditional mean functions. 

Having introduced $\tau(\textbf{x})$, we can summarize the standard definitions of \emph{prognostic} and \emph{predictive} covariates, which are sometimes confused in practice. \cite{Ruberg2015} A baseline covariate $x_j$ is considered prognostic if it is associated with the outcome $Y$, regardless of the treatment administered. Conversely, if the treatment effect $\tau(\textbf{x})$ varies systematically with a covariate $x_j$, it is termed predictive. A covariate can be both prognostic and predictive, meaning it may be associated with $Y$ under the control treatment while also influencing the treatment effect. Additionally, $x_j$ can be non-informative, indicating that it is unrelated to both the outcome and the treatment, yet it may still be measured due to a lack of scientific knowledge at the planning stage.

In observational studies, the treatment assignment probability, known as \emph{propensity score}, is commonly modeled as a function of baseline covariates that dictated actual treatment assignment, $p_{obs}(\textbf{x}_i)=P(A=1|X=\textbf{x}_i)$. Propensity is also an important component of heterogeneity assessment in RCTs, being integral parts of doubly robust (DR) estimators of individual and average treatment effects, to be introduced later in this section. The key advantage of DR estimators is that they are consistent when at least one of the two modeled functions (outcome, $m_a(\textbf{x})$, or propensity, $p(\textbf{x})$) is correct.  In RCTs,  propensity scores are known constants, thus DR etimators are always consistent.  
 
We will now examine ML approaches for estimating $\tau(\textbf{x})$, using $\hat{\tau}(\textbf{x})$ to denote the estimated CATE in contrast to the true population CATE. In recent years, numerous methods have been proposed for this purpose. In this paper, we focus on several meta-learner approaches,\cite{JacobsCATE} including the T-, S-, X-, R- and DR-learners. These techniques build upon existing off-the-shelf supervised learning methods, such as Random Forest and Gradient Boosting. We will also consider a modified machine learning approach known as Causal Forest,\cite{AtheyGRF2019} that predicts CATE, while other examples of similar techniques include Bayesian Causal Forest\cite{HahnBART} and Causal Boosting\cite{Powers2018}. In their study, Lipkovich et al.\cite{ILDS} benchmarked these techniques, along with several others, in terms of their ability to estimate CATE, including the properties of novel subgroups that arise when CATE is dichotomized. Our current benchmarking builds upon this work but focuses on different aspects, specifically the ability to accurately identify biomarkers that drive the systematic variability of $\tau(\textbf{x})$—that is, predictive biomarkers—particularly in the context of using SHAP values.

The representation $\tau(\textbf{x}) = m_1(\textbf{x}) - m_0(\textbf{x})$ indicates a straightforward modeling approach known as the T-learner, where ``T'' stands for ``two''. In this method, the conditional mean functions $m_0(\cdot)$ and $m_1(\cdot)$ are estimated by fitting arm-specific regression models, which entails modeling $Y$ against $\textbf{x}$ separately for the control and active treatment arms. The estimated CATE is then given by
\[ \widehat{\tau}_T(\textbf{x}) = \hat{m}_1(\textbf{x}) - \hat{m}_0(\textbf{x}). \]

The S-learner, where ``S'' stands for ``single'', is another intuitive approach that can be employed. In this method, a single outcome model is constructed by utilizing the entire dataset and fitting $Y$ against $(\textbf{x}, A)$, with the treatment assignment included as an additional covariate.  However, various studies have suggested \cite{Foster2011, Hermansson2021, ILDS} that it is more efficient to enhance the training data $\textbf{W}$ by incorporating pre-computed treatment-covariate interactions before fitting the single model:
\begin{equation}
\textbf{W} = (Y, \textbf{x}, A, \textbf{x} \cdot I_{A=0}, \textbf{x} \cdot I_{A=1}), 
\label{eq:S_learner_W}
\end{equation}
where $I_{B}$ denotes the indicator function whose value is $1$ if condition $B$ holds, and $0$ otherwise. This approach has been shown to yield better properties and is standard practice, as exemplified in the \texttt{rlearner} GitHub R package \cite{rlearner}. This version will be considered in this paper whenever we discuss S-learning. After fitting the regression model $m(\textbf{x}, A)$ to the data $\textbf{W}$, the CATE is estimated as:
\[ \widehat{\tau}_S(\textbf{x}) = \hat{m}(\textbf{x},A=1) - \hat{m}(\textbf{x}, A=0). \]

While the aforementioned methods are somewhat intuitive, later research \cite{Kuenzel2019} has identified inherent issues in their performance. For instance, regarding T-learner approaches, even if each conditional mean function is regularized towards zero separately, there is no guarantee that their difference will also be regularized. In the case of S-learner methods, excessive shrinkage may lead to missed treatment effects. Issues related to varying complexities in the treatment arms have been noted, prompting the development of new approaches, such as X-, R- and DR-learner, intended to address these challenges, albeit with potentially less intuitive frameworks \cite{ILDS}. 

The X-learner\cite{Kuenzel2019} first imputes counterfactual outcomes for each treatment group, and then fits new models to regress the imputed treatment effects against the covariates. Finally, the CATE is estimated as a weighted average, typically using propensity scores to account for the probability of receiving each treatment. Specifically, first estimates of ITE are computed for subjects in the treatment arm by contrasting their observed potential outcomes $Y_i(1)=Y_i, i\in \{i: A_i=1\}$ with the counterfactual outcomes $\widehat{Y}_i(0)$ predicted from the response model $m_0(\textbf{x})$ estimated from the control arm. At the second step, ITE scores $\widehat{\tau}_i=Y_i-\widehat{m}_0(\textbf{x}_i), i \in \{i: A_i=1 \}$ are modeled as a response variable using data from the treatment arm alone, resulting in a CATE estimator $\widehat{\tau}_1(\textbf{x})$ that can be evaluated on any subject as a function of the candidate covariates $\textbf{x}$. Similarly, another CATE  estimator $\widehat{\tau}_0(\textbf{x})$ is constructed utilizing the observed outcomes from subjects in the control arm and their counterfactual responses predicted from a model for $m_1(\textbf{x})$ estimated in the treatment arm, $\widehat{\tau}_i=\widehat{m}_1(\textbf{x}_i)-Y_i$, $i \in \{i: A_i=0\}$. Then $\widehat{\tau}_0(\textbf{x})$ is obtained by regressing $\widehat{\tau}_i$ on the covariates using observations from the control arm. The final estimator for the X-learner is obtained as
\[
\widehat{\tau}_X(\textbf{x})=w(\textbf{x})\widehat{\tau}_0(\textbf{x})+(1-w(\textbf{x}))\widehat{\tau}_1(\textbf{x}), 
\]
where the weight function is often taken as the estimated propensity score, $w(\textbf{x})=\widehat{p}(\textbf{x})$ or a constant probability of treatment assignment (for RCT).

The R-learner is motivated by a natural decomposition proposed by Robinson\cite{Robinson1988}, which involves regressing residuals on residuals while employing cross-fitting \cite{Jacob2020} to mitigate overfitting. Specifically, first the transformed outcomes (or \emph{pseudo-observations}) are calculated:
\begin{equation}\label{eq.TransOutcome}
\widehat{\psi}_{R,i}=\frac{Y_i-\widehat{m}(\textbf{x}_i)}{A_i-\widehat{p}(\textbf{x}_i)}, 
\end{equation}
where $m(\textbf{x})=E(Y|\textbf{x})$ is the overall response function, capturing the main effect of the covariates on the outcomes in the pooled data.
The R-estimator $\widehat{\tau}_R(\textbf{x})$ is then obtained by minimizing a weighted penalized loss with $\widehat{\psi}_R$ as the response variable and subject weights set as $w_i=(A_i-\widehat{p}(\textbf{x}_i))^2$ 

The ideas of the R-learner were incorporated in the recently developed causal inference version of the popular supervised learner Random Forest called Causal Forest \cite{AtheyGRF2019} that introduced several other innovations including recasting Random Forest as a locally weighted non-parametric estimator (averaging treatment effects from terminal nodes across all trees), novel splitting rules maximizing heterogeneity of estimated treatment effects in child nodes, and the idea of ``honest'' estimation of casual estimands by splitting training data into two disjoined sets, one for  finding tree splits and another for estimating local treatment effects.  

Finally, a popular approach to CATE modeling is based on fitting a ML model to a doubly robust learner\cite{Kennedy2020}. Here, first the pseudo-observations are calculated using the augmented inverse probability weighted (AIPW) scores:  
\begin{equation}\label{eq.pseudoOBS}
\widehat{\psi}_{DR,i}=\widehat{m}_1(\textbf{x}_i)-\widehat{m}_0(\textbf{x}_i) + \frac{A_i(Y_i-\widehat{m}_1(\textbf{x}_i))}{\widehat{p}(\textbf{x}_i)}-\frac{(1-A_i)(Y_i-\widehat{m}_0(\textbf{x}_i))}{1-\widehat{p}(\textbf{x}_i)}.
    \end{equation}
Then the pseudo-observations are regressed on the covariates to get the final estimator for CATE,  $\widehat{\tau}_{DR}(\textbf{x}).$

It is worth noting that the AIPW scores are an extension of inverse probability weighted (IPW) scores also known in the literature as \textit{modified outcome}, as they can be thought of as a simple transformation of the original outcome variable. \cite{AtheyImbens2016, Tian2014}
\begin{equation}\label{eq.pseudoOBS}
\widehat{\psi}_{IPW,i}=  Y_i\frac{A_i}{\widehat{p}(\textbf{x}_i)}-Y_i\frac{1-A_i}{1-\widehat{p}(\textbf{x}_i)}=Y_i \frac{A_i-\widehat{p}(\textbf{x}_i)}{\widehat{p}(\textbf{x}_i)(1-\widehat{p}(\textbf{x}_i))}.
    \end{equation}
The modified outcome method (based on regressing the IPW scores on covariates) inspired \textit{modified covariate} and, more generally, \textit{modified loss} methods covering different types of outcomes (with extensions for double robustness) that we will not pursue here.\cite{Tian2014, Chen2017, Huling2021} They share one common element with methods based on pseudo-outcomes, namely, that the estimator of CATE is obtained as a minimizer of an appropriately defined loss function for supervised learning (i.e. regression or classification).  

\begin{table}[h]
    \centering
    \caption{The main notation used in this paper. \label{tab:symbols}}
    \begin{tabular}{@{}ll@{}}
        \toprule
        \textbf{Symbol} & \textbf{Description} \\ \midrule
         $n$ & sample size \\
        $i$ & example/instance/patient index \\ 
        $p$ & dimensionality of covariates \\
        $j,k,l$ & covariate index\\
        $Y, X$ & random variables for the outcome and the vector of covariates \\
        $y, \textbf{x}$ & realization of the outcome and the vector of covariates\\
        $A$ & the treatment indicator\\
        $Y^{(0)}$ and $Y^{(1)}$ & hypothetical outcome under treatment $0$ and $1$ \\ 
        $ \widehat{\psi}_{R/DR}$  & pseudo-observations for R/DR-learner\\
        $\tau(\textbf{x})$ & CATE estimand \\
        $\widehat{\tau}_{S/T/X/R/DR}(\textbf{x})$ & estimated CATE using S/T/X/R/DR-learner \\
        $\mathcal{P} = \{1, \ldots, p\}$ & set of covariates \\ 
        $S$ & subset of covariates, i.e. $S \subseteq \mathcal{P}$ \\ 
        $\mathcal{V}al(S)$ & model output using covariate set $S$ \\
        $ \Delta(j, S)$ & marginal contribution of covariate $j$ for a subset $S$ \\
        $\phi^i_j$ & Shapley value of covariate $j$ on example $i$ (instance-level); index $i$ dropped unless explicitly needed\\
        $\Phi_j$ & Shapley value of covariate $j$ (summary level, i.e. across all instances)\\
     \bottomrule
    \end{tabular}

\end{table}

\subsection{Explainable Machine Learning\label{sec.XML}}
This section provides a general overview of explainable machine learning, emphasizing a widely accepted principle within the machine learning community: \emph{the ability to explain a machine learning (e.g. a complex prediction model's) output is crucial for its users}. Improving model transparency fosters trust in the model's predictions and guarantees safety in using them as the basis of making decisions or taking actions (e.g. optimizing patient treatment). This in turn promotes the ease of adoption of the model by stakeholders (e.g. clinicians, patients, regulators). It can also lead to uncovering new domain area insights (e.g. previously unknown underlying mechanisms of disease or treatment). Finally it can allow us to understand when and how a model fails and how to correct this (e.g. identify bias in the model which can be addressed by improving data collection to obtain a more representative sample of the underlying population).
 
Predictive modeling and machine learning have evolved from the use of simpler models, such as generalized linear models (GLMs), to more complex approaches like ensemble methods and deep neural networks. This shift, primarily motivated by the pursuit of higher predictive accuracy, has resulted in the emergence of intricate ``black-box'' models that are not easily interpretable or subject to human examination, as noted in the literature.\cite{SHAPANALYTICS} To overcome this limitation, a recent trend has been the development of innovative methods to visualize the inner workings of "black-box" models, known as \emph{Explainable Machine Learning (XML)}. The terms \emph{Interpretable Machine Learning (IML)} and \emph{eXplainable Artificial Intelligence (XAI)} are often used interchangeably in the literature. They refer to methods designed to make the predictions or inner workings of ML/AI models more comprehensible to humans, particularly domain experts. Significant research has emerged in recent years in this area.\cite{XML} At its core, XML focuses on both model-specific and model-agnostic approaches that aim to simplify and project key elements of complex, high-dimensional models into lower dimensions. Examples of model-agnostic methods include permutation variable importance, partial dependence plots, Local Interpretable Model-agnostic Explanations (LIME)\cite{LIME}. Model-specific examples include saliency maps\cite{simonyan2014deep}, DeepLIFT\cite{shrikumar2017learning}, Gradient-weighted Class Activation Mapping (GradCAM)\cite{selvaraju2017grad}, SmoothGrad\cite{smilkov2017smoothgrad} and other gradient-based methods commonly used in the context of artificial neural networks. For a summary of the aforementioned and several other approaches, we direct the reader to the comprehensive and accessible book by Molnar (2024)\cite{Molnar}. In this paper we will focus on SHapley Additive Explanations (SHAP), an approach developed in 2017 by Lundberg and Lee\cite{Lundberg2017}. SHAP has both model-agnostic (generally applicable to any model) and model-specific implementations (leveraging specific model structure) and is grounded in the concept of Shapley Values from cooperative game theory, which dates back to 1953\cite{SHAPLEY_ORIGIN}. Due to its flexibility and favorable theoretical properties, the method has rapidly gained popularity within the predictive modeling and machine learning community as a method for fairly assessing the marginal contributions of each predictor to the overall predictive capability of models. The next two subsections will explore the concepts of Shapley Values and SHAPs in turn.

\subsubsection{Shapley values}\label{sec.SHAP_VALUES}
The general theory of Shapley Values was developed to provide a fair method for allocating rewards among a set of players contributing to a specific outcome. In our setting `players' correspond to the model's covariates, the `outcome' corresponds to the model's output (prediction) and `rewards' to assigning credit, or importance to each covariate for its contribution to the model's output. In loose terms, to measure the contribution of any given player (covariate) $j$, one can consider various coalitions of players (covariate subsets) both with and without that player (covariate), tracking how the outcome (prediction) changes when player (covariate) $j$ is excluded. The central idea is to average the contributions across all possible coalitions (subsets of covariates). To formalize this, let $\mathcal{P} = \{1, \ldots, p\}$ represent the set of players (covariates).


For any coalition (subset of covariates) $S \subseteq \mathcal{P}$, we denote the resulting outcome value (model prediction) from their `collaboration' with $\mathcal{V}al(S)$, assuming $\mathcal{V}al(\varnothing)=0$ where $\varnothing$ is the empty set. The idea with Shapley Values is to distribute the total outcome  $\mathcal{V}al(\mathcal{P})$ among the players (covariates) and to this end, notation is needed for the marginal contribution of $j$ for any given coalition $S \in \mathcal{P} \setminus  j$, defined as:
\[ \Delta(j, S) = \mathcal{V}al(S \cup j) - \mathcal{V}al(S).\] The Shapley Value $\phi_j$ then is the average contribution over all such coalitions:
\begin{equation}\label{eq.SV}
    \phi_j = \sum_{S \in \mathcal{P}\setminus  j} \frac{|S|! (p-|S|-1)!}{p!} \Delta(j, S)
\end{equation}
To get a better intuition in the covariate selection setting, the Shapley Value can take the following form:
\begin{equation}
\phi_j = \frac{1}{\text{\# of covariates}} \sum_{\text{all possible subsets excluding covariate } j} \left[ \frac{\text{change in model output due to the addition of covariate } j \text{ to the subset}}{\text{total number of subsets of the given size excluding covariate } j} \right]
\end{equation}

The resulting Shapley Values are known to satisfy certain axioms, which are briefly outlined below. These properties are frequently highlighted as key advantages of using Shapley Values; however, it is important to note that they do not generally hold under the approximations discussed later \cite{Sundararajan}. The \emph{dummy} property (often replaced by the related `\emph{missingness}') implies that covariates absent from a coalition have no contribution to the output. \emph{Local accuracy} implies that the explanatory model's prediction on a given datapoint matches that of the original model. \emph{Additivity} (often replaced by the related `\emph{consistency}') implies that if the model changes such that the marginal contribution of covariate $x_j$ increases or stays the same over all possible subsets of covariates, then the Shapley value for covariate $x_j$ must not decrease.
\begin{enumerate}
\item[(i)] \emph{Local Accuracy (Efficiency)}. The covariate contributions add up to the total: $\sum_{j \in \mathcal{P}} \phi_j =  \mathcal{V}al(\mathcal{P}).$   
\item[(ii)] \emph{Symmetry}. If two covariates $j,k$ contribute equally to the model's prediction, they should have the same reward:\\
If $\mathcal{V}al(S \cup j) = \mathcal{V}al(S \cup k)$ for all $S \subseteq 
 \mathcal{P} \setminus  \{j, k\}$, then $\phi_j=\phi_k$.
\item[(iii)] \emph{Dummy (Null Player)}. No contribution to the model prediction gives zero reward:\\
$\mathcal{V}al(S \cup j) = \mathcal{V}al(S)$ for all $S \subseteq 
 \mathcal{P}$ then $\phi_j=0$.   
\item[(iv)] \emph{Additivity}: For a game with combined contributions $\phi$ and $\phi'$ the respective Shapley Values are $\phi_j + \phi'_j$. 
\end{enumerate}
As noted by Lundberg and Lee \cite{Lundberg2017}, a direct implementation of the expression given in (\ref{eq.SV}) in the context of variable importance scoring for predictive modeling would necessitate refitting the model for each possible subset of covariates. The covariate subsets increase combinatorially with the number of candidate covariates, which poses a significant limitation in practice. It has been observed that the computations become intractable when the number of covariates exceeds ten.\cite{Aas} Consequently, direct implementations are deemed infeasible, prompting various authors to propose alternative methods for estimating Shapley Values. 
A groundbreaking method in this area is the SHapley Additive exPlanations (SHAP) developed by Lundberg and Lee\cite{Lundberg2017}, which will be discussed in the following section. 

\subsubsection{SHapley Additive exPlanations (SHAP)}\label{sec.SHAP_THEORY}

Lundberg and Lee\cite{Lundberg2017} adopted the perspective that any explanation of a model’s prediction can be viewed as a model in its own right, referred to as an `explanation model'. This model serves as an interpretable approximation of the original, arbitrarily complex model. 
They formalized the concept of an interpretable approximation $g(\cdot)$ of the original model $f(\cdot)$, which unifies six previously proposed explanation methods, including LIME\cite{LIME} and DeepLIFT\cite{shrikumar2017learning}. The following section highlights the key aspects of this idea, although a comprehensive theoretical treatment is beyond the scope of this paper.

Consider a given observation $i$ with $\mathbf{x}_i \in \mathbb{R}^p$, representing its covariate vector with $p > 1$ covariates, for which we aim to explain the prediction $f(\mathbf{x}_i)$. Here, $f$ denotes the (presumably complex) prediction model. 
The central idea of SHAP is to explore the solution to (\ref{eq.SV}) within the framework of additive covariate attribution models, which allows us to decompose\cite{Aas} the model prediction $f(\mathbf{x})$ in an additive manner
\begin{equation}\label{eq.AdditiveSH}
f(\mathbf{x}_i) =   \phi^i_0 + \sum_j^p \phi^i_j ,
\end{equation}
where the intercept $\phi^i_0 = E[f(\textbf{x})],$ often termed the \emph{baseline prediction}, can be thought of as the ``leftover'' part of the prediction that is not explained by the covariates’ contributions. 

Expressed in this way, the Shapley Values explain the difference between an individual prediction and the global average prediction. It is important to highlight that this approach provides insights into variable importance at the observation level, which can be particularly valuable in practical applications. Crucially, it isolates the contribution of a given covariate assuming the specific value it has on a given observation, to the model's output. 
There exists a unique solution to (\ref{eq.AdditiveSH}) that satisfies the properties listed in Section \ref{sec.SHAP_VALUES} \cite{Lundberg2017}.
While a thorough discussion of these properties is beyond the scope of this paper, it is worth noting that they represent natural requirements from a human perspective which are not universally satisfied by all types of explanatory models (i.e. model explainability methods).


\subsubsection{Instance-level vs. Summary-level SHAP values \label{sec.SummarySHAP}}

As elaborated on earlier, SHAPs offer insights regarding how each covariate's value is contributing to a single prediction. However, often a summary measure of variable importance is required which can be easily obtained from the individual-level contributions. To clarify matters, we emphasize the difference between \emph{instance-level} and \emph{summary-level} SHAPs\footnote{In the machine learning literature \emph{instance-level} explanations are commonly referred to as \emph{local} explanations and \emph{summary-level} explanations as \emph{global} explanations.}.
\begin{description}
\item[Instance-level importance] refers to the importance of a specific covariate $x_j$ for a particular instance (observation), $\phi^i_j$, capturing how much the covariate's particular value (e.g. $\text{patient age} = 72$) contributes to the prediction of outcome for that specific instance.
    
\item[Summary-level importance] refers to the overall importance of a covariate across the entire dataset or population (or a select subset of it). It is calculated considering the average impact of each covariate across all instances. For a given covariate $x_j$ ($1 \leq j \leq p$) the global SHAP importance $\Phi_j$ is defined as follows:
\begin{equation}\label{eq.SummarySHAPfromInstances}
\Phi_j = \frac{1}{n} \sum_{i=1}^n |\phi^{i}_{j}|.
\end{equation}
\end{description}

The importance measure of (\ref{eq.SummarySHAPfromInstances}) quantifies the average absolute contribution of covariate $x_j$ across all observations in the dataset, providing a clear indication of how critical $x_j$ is in driving the model’s predictions. This overall measure has become a standard in data science for ranking covariates, exemplified by the widely used SHAP summary plot available in various software packages, including R and Python.\cite{Molnar} 
In the summary plot, we observe initial indications of the relationship between a covariate's value and its impact on the prediction. However, to understand the exact nature of this relationship, we need to examine SHAP dependence plots. \cite{Molnar} These plots can be seen as the widely used partial dependence plots, introduced by Friedman\cite{friedman2001greedy} in the context of gradient boosting, that aim to show the marginal effect that a covariate has on the predicted outcome of a model. 


\subsubsection{Model-agnostic vs. model-specific SHAP estimators\label{sec.Kernel}}

An approach for approximating SHAP values $\phi_j$, $j=0, \dots, p$ that is agnostic to the underlying model was introduced by Lundberg and Lee \cite{Lundberg2017}, and further refined by Aas et al. \cite{Aas}. This model-agnostic `kernel' method (more on its name later) allows for the estimation of covariate contributions without relying on specific details of the model, making it versatile and applicable across various machine learning frameworks. Thus, KernelSHAP, is a natural choice for CATE models, which can be multi-stage and may not conform to conventional model classes. 
The KernelSHAP estimator shares similarities with LIME (Local Interpretable Model-agnostic Explanations)\cite{LIME}.  The idea behind LIME is to locally approximate the input-output mapping of an arbitrarily complex black box model $f(\cdot)$ with a more interpretable (e.g. a tree or linear model) `explanation model' $g(\cdot)$. To obtain a local explanation of the original model $f(\cdot)$ on a given instance $\mathbf{x}$ (i.e. to measure the contribution of each covariate to the prediction  $f(\mathbf{x})$), LIME first calculates the predictions of the black box model $f(\cdot)$ on perturbations of point $\mathbf{x}$ (i.e. a set of new points obtained by adding noise to $\mathbf{x}$). The next step is to train a weighted, interpretable model $g(\cdot)$ on the set of perturbed points weighted according to their proximity to $\mathbf{x}$. We can then explain the prediction of $f(\cdot)$ by interpreting $g(\cdot)$. For example, if $g(\cdot)$ is a linear model, we can use its coefficients to determine the contribution of each covariate to the prediction.

KernelSHAP first samples a subset of covariate coalitions (combinations) and calculates the model-based prediction $f(\cdot)$ for each coalition. It then assigns a weight to each such coalition which is similar to weighting datapoints in LIME. However, unlike LIME,  KernelSHAP weights the sampled instances according to the weight the coalition would get in the Shapley value estimation of (\ref{eq.SV}), resulting in  the largest weights assigned to small and large coalitions. The intuition behind this is that we learn most about the importance of individual covariates if we study their effects in isolation. Thus, we extract most information about covariate $x$ when a coalition consists of a single covariate $x$ or if a coalition consists of all covariates but $x$. In the latter case we can learn about the total effect of $x$ including its main effect and interactions with other covariates. If a coalition contains half of all covariates, we learn little about any individual covariate’s contribution, as there are many such coalitions with half of the covariates. The function that assigns weights to each datapoint is the eponymic `kernel' function of the method. Finally, KernelSHAP fits a weighted linear model and returns its coefficients as the estimates of the Shapley values. While there are many alternatives for the loss, penalty, and kernel function used in fitting the weighted linear model, Lundberg and Lee\cite{Lundberg2017} presented a specific combination that guarantees the desired properties of local accuracy, missingness and consistency, as shown in their Theorem 2. This combination serves as the foundation of the KernelSHAP approach. As the method relies on sampling covariate combinations, it should be obvious that it does not produce \emph{exact} Shapley values, but rather their \emph{estimates}. Moreover, as it requires fitting a new model $g(\cdot)$ on each datapoint and sampling several covariate coalitions to guarantee a good estimation, it faces limitations due to computational constraints, as we will demonstrate later. 


In addition to KernelSHAP, several model-specific SHAP approximations have been proposed for cases where the function $f(\cdot)$ is restricted to specific model classes. These model-specific methods, which include LinearSHAP (for linear models), TreeSHAP (for tree-based models and ensembles thereof), DeepSHAP and GradientSHAP (both for neural network models), leverage the particular structure and properties of the models they target, allowing for more efficient and accurate computation of SHAP values. Of particular importance in this paper is the TreeSHAP approach,\cite{lundberg2020local} which applies specifically to tree-based models, such as decision trees, random forests, and gradient boosting models like XGBoost. TreeSHAP takes advantage of the hierarchical structure of tree ensembles, enabling it to compute exact SHAP values in polynomial time, which is substantially more efficient than the general SHAP algorithm having exponential time complexity. More specifically, for estimating the SHAP value of a single covariate,  TreeSHAP has a computational complexity of $O(tld^2)$, where
$t$ is the number of trees in the ensemble $l$ the maximum number of tree leaves and $d$ the maximum tree depth. In other words, its running time depends quadratically on the maximum tree depth, a user determined parameter of the model. On the other hand, KernelSHAP has a complexity of $O(n_c)$ for calculating the SHAP value for a single covariate, where $n_c$ is the number of covariate combinations examined. For small numbers of covariates $p$, i.e. $>8-15$, depending on the implementation, calculation of exact SHAP values is feasible (the computational complexity is $O(2^p)$ as the possible combinations of $p$ covariates are $2^p$). For larger values of $p$ KernelSHAP approximates the SHAP value by subsampling $n_c$ out of the $2^p$ possible covariate combinations. The smaller $n_c$ is, the worse the approximation but the faster the estimation of the SHAP values. The different strategies for sampling the space of coalitions have been the subject of recent study \cite{olsen2024improving}. In short, the running time of KernelSHAP depends exponentially on the number of covariates in the data for a good approximation of SHAP values. This makes TreeSHAP much faster (and exact) in high dimensional settings. By exploiting the tree structure, TreeSHAP can precisely attribute the contribution to the model's prediction among the covariates, providing consistent and accurate variable importances.  Overall, the development of model-specific SHAP methods has significantly advanced the usability and effectiveness of SHAP values in practical machine learning applications, particularly when dealing with complex model structures such as tree ensembles and large datasets. For a recent summary of several model-agnostic and model-specific methods for estimating SHAP values, we direct the reader to  Chen et al. \cite{chen2023algorithms}.

The development of SHAP methodologies has led to the creation of several packages catering to different requirements. The \texttt{shapr} package, available on CRAN, extends the KernelSHAP approach to handle dependencies between covariates, making it a robust tool for general model interpretations. Another notable tool is \texttt{ShapleyR}, accessible on GitHub, which provides utilities for calculating Shapley values and integrates well with a variety of models. Additionally, the \texttt{XML} package offers a comprehensive framework for model explanation and interpretation, supporting the implementation of SHAP values alongside other interpretive methods. For R users, \texttt{shapper} and \texttt{kernelshap} serve as CRAN ports of the popular Python package \texttt{shap}, simplifying the computation of SHAP values within the R ecosystem.

For model-specific needs, the \texttt{SHAPfor} package focuses exclusively on XGBoost models. By leveraging the unique structure of tree-based ensembles, this package ensures efficient and accurate computation of SHAP values, enhancing the interpretability of XGBoost predictions. These packages collectively enhance the usability and applicability of SHAP values across various machine learning models and frameworks, offering both model-agnostic and model-specific solutions to suit diverse use cases.

\subsection{Explainable causal machine learning \label{sec.SHAP_CATE}}

Having explored the fundamentals of Shapley values and SHAP in the context of supervised learning, we now shift our attention to their application in causal inference models aimed at estimating CATE. This transition introduces some complexity, as CATE is typically derived from the combination of multiple models. For example, in the T-learner approach, the CATE predictions are calculated as the difference between the predictions from individual arm-specific outcome models. This approach places CATE outside the realm of standard machine learning model classes.

\subsubsection{SHAP based on modified ML models for estimating CATE}
We start with the methods that modify existing `off-the-shelf' machine learning models  into targeting CATE instead of predicting outcomes; these include Causal Forest\cite{AtheyWager2019}, Causal Boosting\cite{Powers2018}, Causal MARS\cite{Powers2018}, Bayesian Causal Forest\cite{HahnBARTCRAN}, and Causal BART\cite{HahnBART}. In these methods, there is a direct model mapping of the covariates to CATE: $\texttt{M}: \textbf{x} \rightarrow \tau$, which can be explained through SHAP. To this end, we can use a model-agnostic procedure, such as \texttt{KernelSHAP}\cite{KernelSHAPpackage}. For methods that build upon decision trees, such as the Causal Forest, it is also possible to explore the derivation of model-specific SHAP, in this case, TreeSHAP. Unfortunately, there are no packages that implement \texttt{TreeSHAP} that support models generated by the \texttt{grf}\cite{GRF} package, which implements Causal Forest. For all the methods mentioned above, there are no off-the-shelf implementations of model-specific SHAP, making the usage of model-agnostic SHAP, through \texttt{KernelSHAP}, the only feasible solution.

\subsubsection{SHAP based on meta-learners for estimating CATE}
Typically, meta-learners estimate CATE by building and combining off-the-shelf machine learning models in various ways. Consequently, deriving SHAP explanations can be challenging, particularly for model-specific SHAP. For our purposes—deriving SHAP values from CATE models—it is useful to categorize meta-learners into two main categories:

\begin{description}
    \item[Reducible meta-learners:] These methods reformulate the estimation of CATE  as a predictive supervised machine learning problem, possibly involving the estimation of nuisance functions. CATE is ultimately derived as an outcome of supervised learning with an appropriate loss function. 
    This category includes methods such as the R-learner, and DR-learner, as well as modified outcome methods \cite{Tian2014}. For methods in this category, SHAP values can be derived directly from the final model. For the DR-learner, the CATE estimate involves regressing the pseudo-observations defined in equation (\ref{eq.pseudoOBS}) against the baseline data: $\widehat{\psi}_{DR} \sim \textbf{x}$, allowing us to derive SHAP values from this mapping. Similarly, for R-learning, the transformed outcomes are regressed against the baseline covariates via XGBoost with a modified loss (see equation (\ref{eq.TransOutcome})): $\widehat{\psi}_{R} \sim \textbf{x}$, and SHAP values can again be derived from this mapping. In both cases, it is straightforward to use either model-agnostic or model-specific explanations, depending on the model used in the final mapping. Furthermore, the modified loss proposed by \cite{Chen2017} also falls in this category of methods. Here, a machine learning regression model is used to minimize a loss to construct so-called \emph{benefit scores}, which represent a transformation of CATE.

     \item[Irreducible meta-learners:] These meta-learners cannot be reduced to a supervised learning problem. They involve multiple stages and models combined to estimate CATE, and unlike the reducible meta-learners, there is not a single model that minimizes a loss function to derive CATE. This category includes methods such as S-, T- and X-learner. Similar to modified machine learning methods, for these CATE estimators, it is relatively straightforward to derive explanations through model-agnostic approaches, such as \texttt{KernelSHAP}. On the other hand, there are no off-the-shelf methods for deriving model-specific explanations like \texttt{TreeSHAP} for these learners. While heuristic approaches can be employed to use model-specific explanations, this may not be straightforward. For instance, in the modified S-learner, where the input space is expanded with pre-computed treatment-covariate interactions (more details in Section \ref{sec.CATE} on how this strategy improved performance), it remains uncertain whether the meta-learner continues to represent a mapping from the covariate space to $\mathbb{R}$, thus introducing uncertainty in deriving direct SHAP values. Conversely, for the T-learner, combining SHAP values of individual models to obtain the corresponding SHAP value of the linear combination of the models (i.e., the output of CATE estimators) suggests itself due to the additive property of SHAP; however it will tend to overemphasize the prognostic covariates. As mentioned, there are no off-the-shelf solutions currently available; therefore, our recommendation is to use these learners with KernelSHAP, while model-specific explanation methods can be considered an area requiring further research.
\end{description}

\subsection{Strategies for deriving SHAP explanations from CATE estimators}\label{sec.strategies}
Building on the discussions from the previous two sections, we will now summarize our recommendations for deriving SHAP explanations of CATE methods. We outline three strategies: the first two are inspired by our earlier discussions and are tailored to specific CATE methods, while the third is a more generic and pragmatic approach.

\begin{description}
    \item[Strategy 1 (tailored to irreducible methods):] For modified machine learning models and irreducible meta-learners, one option is to derive SHAP values directly for the mapping  $\textbf{x} \rightarrow \hat{\tau}.$ As discussed, this option is available regardless of the nature of the CATE modeling strategy. However, only model-agnostic SHAP values can be derived directly using the capabilities of current packages. This can be achieved in R packages such as \texttt{shapr}\cite{SHAPR} and \texttt{KernelSHAP}\cite{KernelSHAPpackage} by using user-defined custom functions. For example, in the \texttt{KernelSHAP} package, the \texttt{predict()} function needs to be replaced by a custom function, as illustrated in our Supplementary material. The main point is to set up a function that returns the predicted CATE for any covariate input.

    \item[Strategy 2 (tailored to reducible methods):] Another strategy that is tailored to reducible meta-learners is to derive the SHAP values directly from this final supervised model that estimates CATE. In this case, the type of the regression model determines what type of SHAPs estimator can be applied: if the model is tree-based, TreeSHAP suggests itself due to being optimized for such modeling classes (provides exact SHAP values efficiently). However, in principle, it is still possible to consider deriving the model-agnostic SHAP, such as \texttt{KernelSHAP}. This might be the natural choice when the regression model is of quite different nature, such as relying of model averaging of very different model types e.g., Superlearner\cite{SuperLearner}. 

    \item[Strategy 3 (using a surrogate model; applicable to all methods).]
    The last strategy, which this paper explores, is a pragmatic method that overcomes the aforementioned implementation challenges by fitting an additional (surrogate) model, $\texttt{M}^{(2)}$, to regress the estimated CATE, $\widehat{\tau}(\textbf{x})$, against the baseline predictors $\textbf{x}=(x_1, ..., x_p)$, where $p$ represents the total number of covariates. The superscript ``(2)'' signifies this extra modeling step. Subsequently, SHAP values are derived from $\texttt{M}^{(2)}$ using the standard machinery. 
    The fundamental idea behind this approach is to explain a well-fitted model that represents the estimated CATE as a function of covariates. This pragmatic solution draws parallels with the computation methodology for extracting insights regarding novel subgroups in the precursor to the meta-learner, Virtual Twins.\cite{Foster2011} This idea has been further generalized for generating variable importance by recommending the fitting of any tree ensemble to derive variable importance.\cite{BOOKBIOMARKER} This approach is quite generic, and  can be used with any method that returns an estimate of the CATE, since we calculate the SHAP values using this estimate as an outcome variable. The idea of training a simpler surrogate model to emulate the input-output mapping of a more complex one before using SHAP to explain it has been previously explored by Messalas et al.\cite{messalas2019model}, but it has not -to our knowledge- previously been applied in the context of causal inference. 
\end{description}

As mentioned in Section \ref{sec.SHAP_THEORY}, there is a plethora of SHAP methods that we can utilize. The most popular choices remain KernelSHAP and TreeSHAP. Other implementations exist, including DeepSHAP and GradientSHAP, which are optimized for neural networks (consider Mosca et al.\cite{Mosca} for a full review of existing packages). In our work, we focus on the two most popular choices and how they combine with the CATE strategies to limit the number of possibilities.

Furthermore, whenever we need to build a machine learning model, we use XGBoost. We choose this model because it has demonstrated state-of-the-art performance in tabular data.\cite{SHWARTZZIV202284} We conducted an extensive hyperparameter tuning process to optimize the model's performance. Additionally, we explored various parameter grids to ensure the robustness of our results across different configurations, and our findings confirmed this robustness.

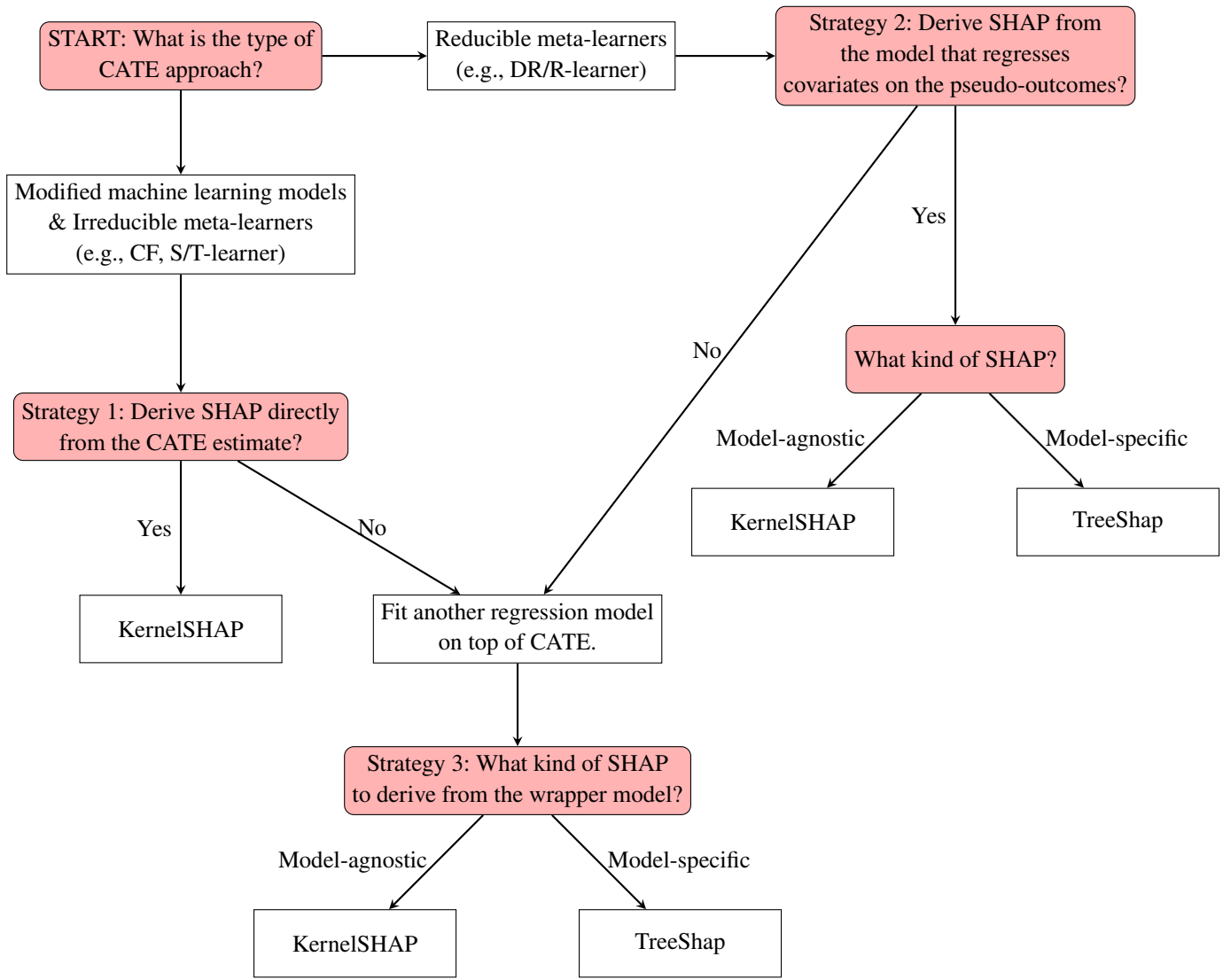
\begin{figure}[htb!]
    \centering
\begin{tikzpicture}[node distance=2cm]
\node (start) [question,align=center] {START:  What is the type of \\ CATE  approach?};
\node (condmean) [process,align=center, below of=start, yshift=-0.5cm] {Modified machine learning models\\ \& Irreducible meta-learners\\ (e.g., CF, S/T-learner)};
\node (pseudo) [process, align=center, right of=start, xshift=3.5cm] {Reducible meta-learners \\ (e.g., DR/R-learner)};
\node (condmean_shap) [question, align=center, below of=condmean, yshift=-1cm] {Strategy 1: Derive SHAP directly \\ from the CATE estimate?};

\node (condmean_shap_kernel) [process, 
align=center, below of=condmean_shap,yshift=-1cm] {KernelSHAP};

\node (condmean_fitreg) [process, right of=condmean_shap_kernel,align=center, xshift=3cm] {Fit another regression model\\ on top of CATE.};
\node (condmean_fitreg_SHAP) [question, below of=condmean_fitreg,align=center, yshift=-0.25cm] {Strategy 3: What kind of SHAP\\ to derive from the wrapper model?};
\node (condmean_fitreg_modelagnostic) [process, align=center, below left of=condmean_fitreg_SHAP, xshift=-1cm, yshift = -1cm] {KernelSHAP};
\node (condmean_fitreg_modelspecific) [process, align=center, below right of=condmean_fitreg_SHAP, xshift=1cm,yshift = -1cm] {TreeShap};

\node (pseudo_CATE) [question, align=center, right of=pseudo, xshift=4cm] {Strategy 2: Derive SHAP  from\\ the model that regresses\\ covariates on the pseudo-outcomes?};

\node (pseudo_CATE_direct) [question, align=center, below of=pseudo_CATE, yshift=-2.5cm] {What kind of SHAP?};

\node (pseudo_CATE_direct_modelagnostic) [process, align=center, below left of=pseudo_CATE_direct, xshift=-1cm, yshift = -1cm] {KernelSHAP};
\node (pseudo_CATE_direct_modelspecific) [process, align=center, below right of=pseudo_CATE_direct, xshift=1cm, yshift = -1cm] {TreeShap};

\draw [arrow] (start) -- (condmean);
\draw [arrow] (start) -- (pseudo);
\draw [arrow] (condmean) -- (condmean_shap);
\draw [arrow] (condmean_shap) -- node[anchor=east] {Yes} (condmean_shap_kernel);
\draw [arrow] (condmean_shap) -- node[anchor=west] {No} (condmean_fitreg);

\draw [arrow] (condmean_fitreg_SHAP) -- node[anchor=east] 
 {Model-agnostic}(condmean_fitreg_modelagnostic);
\draw [arrow] (condmean_fitreg_SHAP) -- node[anchor=west] 
 {Model-specific}(condmean_fitreg_modelspecific);
 
\draw [arrow] (condmean_fitreg) -- node[anchor=east]{}(condmean_fitreg_SHAP);

\draw [arrow] (pseudo) -- (pseudo_CATE);

\draw [arrow] (pseudo_CATE) -- node[anchor=east]{Yes}(pseudo_CATE_direct);

\draw [arrow] (pseudo_CATE_direct) -- node[anchor=east] 
 {Model-agnostic}(pseudo_CATE_direct_modelagnostic);
\draw [arrow] (pseudo_CATE_direct) -- node[anchor=west] 
 {Model-specific}(pseudo_CATE_direct_modelspecific);

\draw [arrow] (pseudo_CATE) -- node[anchor=east]{No}(condmean_fitreg);

\end{tikzpicture}

    \caption{A roadmap illustrating the various principled approaches to derive SHAP values for CATE models. This flow chart provides a clear guide through the different methods of CATE estimation and stages of SHAP implementation.}
    \label{fig:roadmap}
\end{figure}

To sum up, this section illustrated how to derive SHAP values for CATE models. The two major aspects impacting the analysis are (i) the different modeling strategies for estimating CATE and (ii) the various implementations for deriving SHAP values (agnostic or model-specific). Despite the numerous possible combinations of options, this has not yet been clearly summarized in any source. To facilitate understanding, we present a schematic flow chart in Figure \ref{fig:roadmap}, allowing easier navigation across the different options.

\section{Simulation  illustrations}\label{sec.examples}
In this section, we illustrate various aspects of SHAP inference, beginning with standard setting of predictive modeling and then contrasting them with scenarios where the ML model functions as a CATE estimator. It’s important to note that this section is not a comprehensive simulation study—that will be addressed in the following section. Instead, our goal here is to highlight key characteristics of the methods and demonstrate why the model-specific SHAP  is more feasible in practice. These illustrations also serve to motivate the evaluation metrics we consider, providing context for the later simulation benchmarking discussed in Section \ref{sec.LSB}.

\subsection{SHAP for supervised learning\label{sec.supervisedlearningSHAP}}
The first example illustrates how SHAP values can provide insights into a supervised learning model in a straightforward manner. We simulate $400$ observations from the model defined by $y = sin(\pi\!\cdot\!x_1) + \epsilon,$ where $\epsilon \sim N(0,1)$ and $x_1 \sim N(0,1)$. To introduce a bit more complexity, we also generate additional covariates independently as  $x_j \sim N(0,1)$ for $j=2,...,50$. This setup allows the subsequent modeling process to potentially misidentify non-informative covariates as relevant. Next, an XGBoost model, widely recognized as a powerful machine learning method for tabular data\cite{grinsztajn2022why}, was fitted to the dataset. For optimizing hyperparameters, a CV grid search was performed over $\eta, \gamma$, \texttt{Tr.depth}, \texttt{CSamp}, \texttt{SubSamp}, \texttt{ChildW}, and ensemble \texttt{size} $\leq 1000$ using the R package \texttt{xgboost}.\cite{xgboostpackage} 

\begin{figure}[htb!]
    \centering
    \begin{subfigure}[b]{0.35\textwidth}
        \centering
        \includegraphics[width=\textwidth]{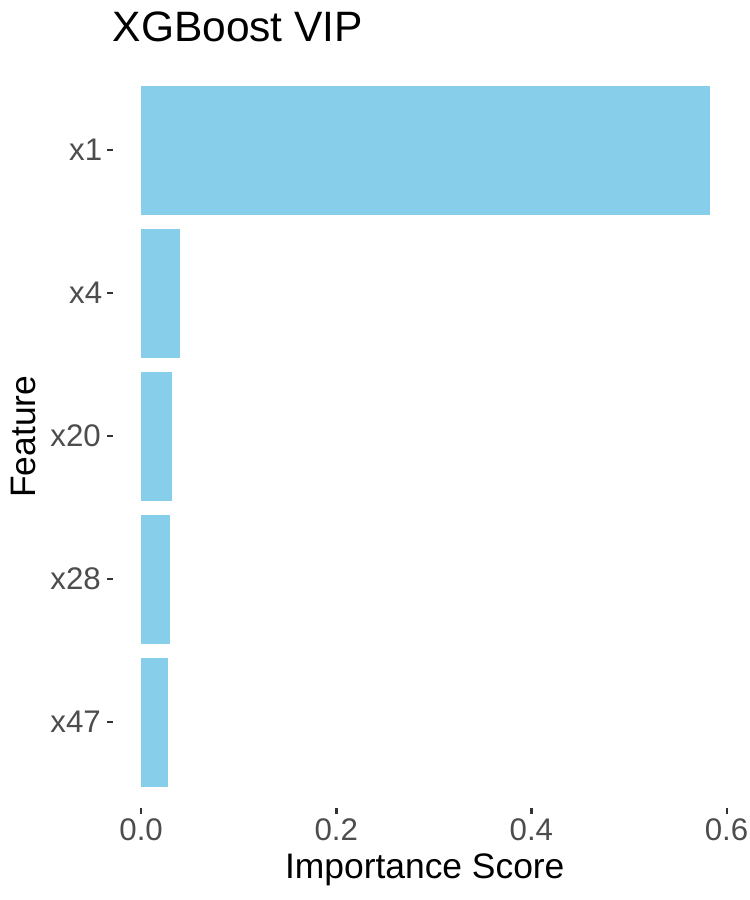}
        \caption{Classical VIP}
        \label{fig:vip}
    \end{subfigure}
    \begin{subfigure}[b]{0.35\textwidth}
        \centering
        \includegraphics[width=\textwidth]{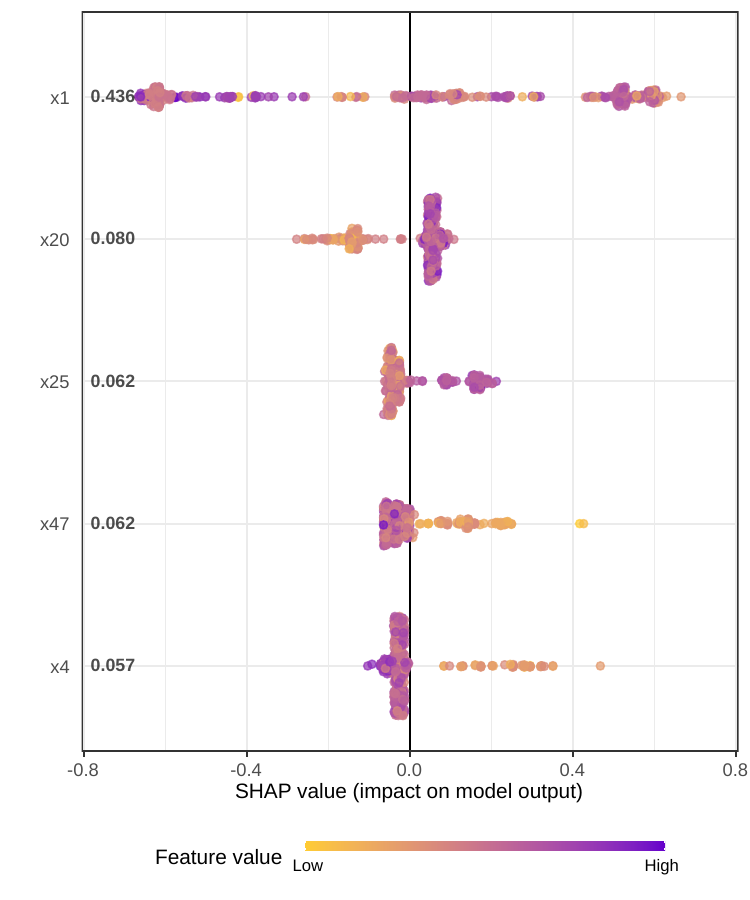}
        \caption{SHAP summary plot}
        \label{fig:shap_summary}
    \end{subfigure}\\
        \begin{subfigure}[b]{0.6\textwidth}
        \centering
        \includegraphics[width=\textwidth]{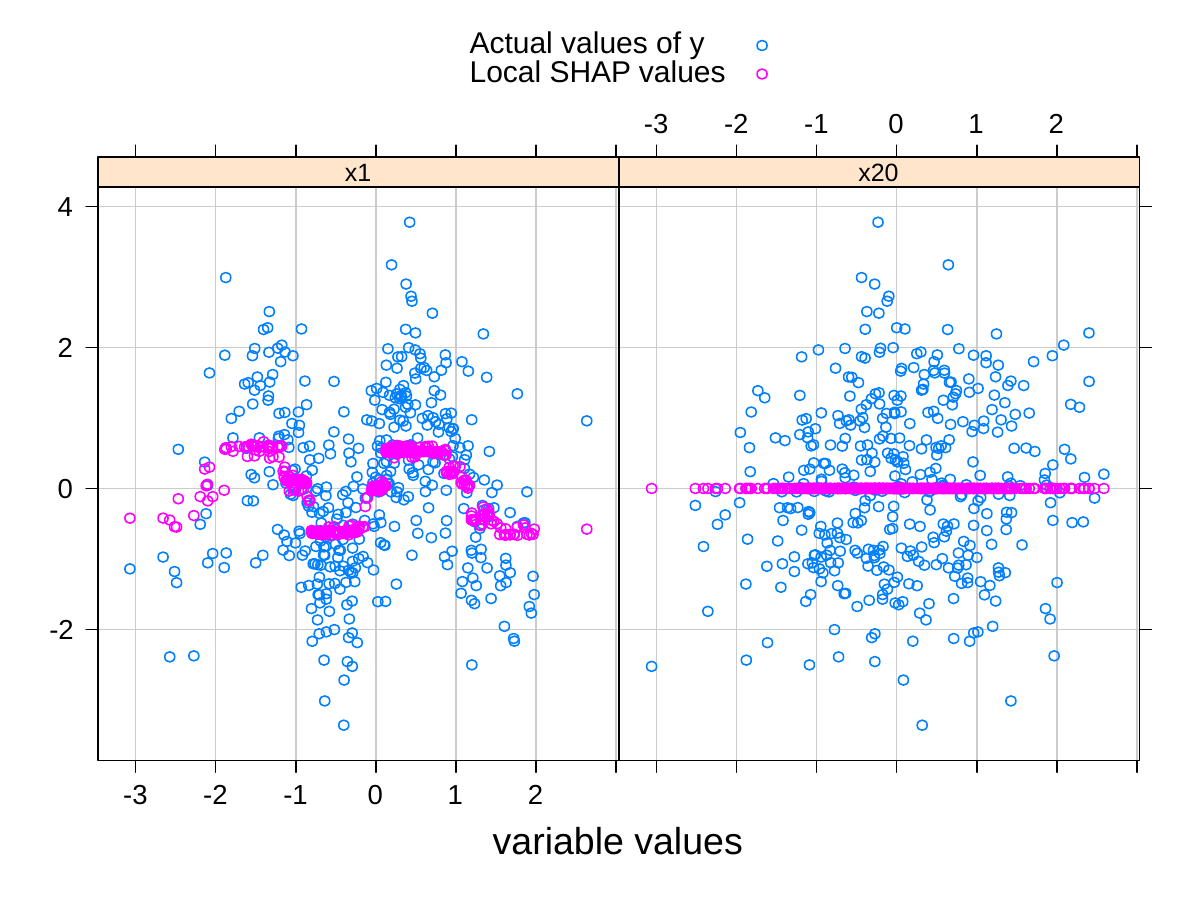}
        \caption{SHAP dependency plot for covariates $x_1$ and $x_{20}$}
        \label{fig:shap_dependency}
    \end{subfigure}
    \caption{Toy example where $x_1$ is the only covariate truly influencing the outcome. (a) illustrates classical covariate importance scores, (b) SHAP summary plots, (c) SHAP dependency plots for two covariates.}
    \label{fig:SinusTrue}
\end{figure}

To better illustrate the potential of SHAP values (specifically tree-SHAPs in this example), we derive classical gain-based Variable Importance measures (VIP) for each $x$-covariate  (this is an impurity based importance). In Figure \ref{fig:vip} we visualize the classical VIP, and given the simplicity of the example, it is not surprising that it readily identifies the truly important covariate, $x_1.$  However,  this alone does not provide any additional insight into the impact of the covariate on the prediction of the model. 

Figure \ref{fig:shap_summary} presents a summary plot of SHAP values, offering a detailed overview of the impact of covariates on the model's output. Each dot represents an observation, with the x-axis showing SHAP values and the y-axis listing covariates, with $x_1$ at the top due to its high importance. Dots are color-coded based on covariate values, indicating their contribution to the prediction. Dots to the right of the zero line show a positive contribution, while dots to the left show a negative contribution. The spread of dots along the x-axis for $x_1$ illustrates the variability of its impact on predictions. In the summary plot, the number associated with each covariate represents the average absolute SHAP value for that covariate. This number quantifies the overall importance of the covariate in the model. A higher value indicates that the covariate has a greater impact on the model's predictions, while a lower value suggests a lesser impact. 

Furthermore, SHAP values can indicate functional relationship between model prediction and covariates. For example, SHAP dependence plots in Figure \ref{fig:shap_dependency} visualize the raw data (blue dots) and the corresponding SHAP values (purple dots) for the covariates $x_1$ and $x_{20},$ which are the top-2 covariates in the SHAP summary plot. The leftmost panel indicates that only $x_1$ significantly contributes to the boosting model’s predictive capability, shifting the predictions away from the average in a pattern reminiscent of a sine function. This effectively uncovers the true underlying functional relationship and highlights another benefit of SHAP values: making it easier to explore and understand underlying patterns in real datasets.  While monotonic trends may be more plausible in biological or medical contexts, it is important to consider the potential for false discoveries in practice. In this example, the SHAP values for the other candidate predictors did not reveal any particular trend, but this may not be the case in real-world or RCT analyses. This is a crucial aspect that we will explore further.  

\subsection{Model-agnostic vs. model-specific SHAP for T-learner \label{sec.ExDirIndirT}} 
The following simulation compares the model-agnostic (Strategy 1) and model-specific SHAP approaches (Strategy 3) discussed in Section \ref{sec.strategies} when applied to CATE estimates derived from a T-learner framework. The data was simulated based on the model $y=-1\!+\!x_1\!+\!x_2 +\!2\!\cdot\!I(x_3>0)\cdot\!A + \epsilon$, where $\epsilon \sim N(0, 0.1)$ and treatment $A \sim$ Bernoulli($0.5$), i.e., 1:1 randomization is assumed. The number of observations is $n=600$ and the number of covariates is $5$: $x_j \sim N(0,1)$ i.i.d., $j=1,...,5$, i.e, two covariates are noninformative. The $x_1$ and $x_2$ are main effects (i.e., prognostic) while $x_3$ is the true treatment modifier (i.e., predictive). The cut-off value for the predictive covariate is zero, leading to a better treatment effect for patients with positive $x_3$ (assuming larger outcomes are beneficial). Covariates $x_4$ and $x_5$ are non-informative and added to make the analysis more challenging. The T-learner model used  RandomForest\cite{RandomForest} as the base learner (with $1000$ trees).

\begin{figure}
    \centering
    \includegraphics[scale=0.8]{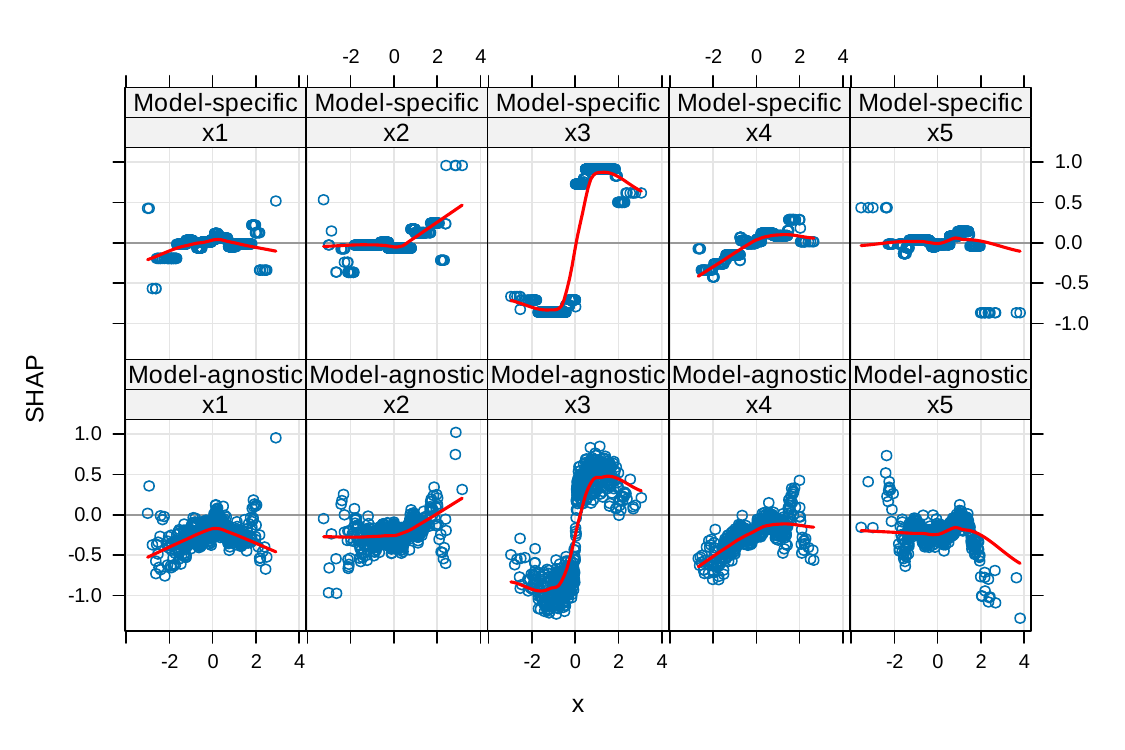}
    \caption{Illustration of the model-agnostic and model-specific SHAP approach for a T-learner CATE model. The only truly predictive covariate was $x_3$ in this example. The resulting SHAPs capture this, ``explaining'' how $x_3$ is pushing the prediction of CATE away from the overall prediction. Although the other covariates are not truly predictive, the fitted CATE model may not have captured that perfectly and as a consequence the predictions depend slightly on $x_1, x_2, x_4$ and $x_5$ too, as is evident from their slight non-zero SHAPs and hints of small trends. In this example, the model-specific approach arguably renders a more clear-cut result and more in line with oracle knowledge.}
    \label{fig.ModelAgnosticVSModelSpecific}
\end{figure}

Focusing first on the model-agnostic SHAP approach, it was implemented using the \texttt{shapr}\cite{SHAPR} R package, which provides model-agnostic KernelSHAP estimation. (Some attempts were also made using the \texttt{kernelshap}\cite{KernelSHAPpackage} package, see further below). By default, this R package accepts certain classes of prediction models directly (such as \texttt{glm}, \texttt{gam}, and \texttt{ranger}), as well as so-called custom models. However, the latter requires additional user specification, since the T-learner is not one of the allowable models in the \texttt{shapr} package. 
In the lower row of Figure \ref{fig.ModelAgnosticVSModelSpecific}, we see the resulting values plotted for each of the five biomarkers. The package presents results according to expression (\ref{eq.AdditiveSH}), which involves splitting the SHAP values into covariate-specific components and an intercept term, $\phi_0$. This intercept was added to the covariate-specific SHAP values before plotting the results in the lower panel. In contrast, the upper panel displays a model-specific approach, using the TreeSHAP estimation, derived from an XGBoost model. The results in the upper panel show less variability compared to those from the kernel approach. Additionally, for the non-predictive covariates ($x_1, x_2, x_4, x_5$), there is a larger deviation from 0 on average.

The corresponding Summary SHAP ranking is presented in Table \ref{tab.SHAPEX} for both approaches. These summary values confirm that the rankings are nearly identical, with $x_3$ being prominent in both cases and lower values assigned to the non-predictive covariates.

\begin{table}[ht]
\caption{\emph{Summary SHAP for the simulation example, for the two types of approaches considered. Both rendered similar ranking, and correctly top-ranked the only predictive covariate ($x_3$)}\label{tab.SHAPEX}}
\centering
\begin{tabular}{cccc}
  \hline
approach & summary SHAP importance & biomarker \\ 
  \hline
\multirow{5}{*}{model-agnostic} & 0.236 & x1 \\ 
 & 0.243 & x2 \\ 
 & \textbf{0.653} & x3 \\ 
 & 0.242 & x4 \\ 
 & 0.243 & x5 \\ 
    \hline
\multirow{5}{*}{model-specific} & 0.037 & x1 \\ 
 & 0.069 & x2 \\ 
 & \textbf{0.792}  & x3 \\ 
 & 0.098 & x4 \\ 
 & 0.051 & x5 \\ 
   \hline
\end{tabular}
\end{table} 

Figure \ref{fig.ModelAgnosticVSModelSpecific} demonstrates that both types of SHAP analyses are qualitatively in agreement, with $x_3$ standing out due to its clear systematic pattern. The SHAP values for the other covariates are noisier, as evident in the graph, and do not reveal any clear trends. However, cherry-picking patterns poses a potential risk in practical situations. Without knowledge of the simulation model, one might be led to believe that, for example, covariate $x_4$ exhibits a slight monotonic pattern, suggesting it is weakly predictive. We include this example intentionally to illustrate that results are rarely perfect with finite data, and decision-making issues related to multiplicity and false discovery rates are relevant in practice. The key point here is that SHAP values provide a starting point for selecting potentially predictive biomarkers. It is logically possible that none of the covariates inspected are truly predictive, thus picking a biomarker with the highest ranking  may not always lead to valid selection. A comprehensive assessment of biomarkers extends beyond merely inspecting rankings and may require thresholding SHAP values; these additional considerations are outside the scope of this paper investigating the performance of SHAP values under various modeling strategies, such as the choice of CATE learner.

This data set was also analyzed using the \texttt{kernelshap} R package. The results were qualitatively similar to the above, albeit noisier than the tree-based solution showing a less compelling discrimination between $x_3$ and the non-predictive covariates. The exact algorithm was used (possible when the number of covariates is $\leq 8$) and the run took $18047$ seconds on a linux platform (R 4.0.2, 14 GB RAM, 16 CPUs). 

\begin{figure}
    \centering
    \includegraphics[scale=0.8]{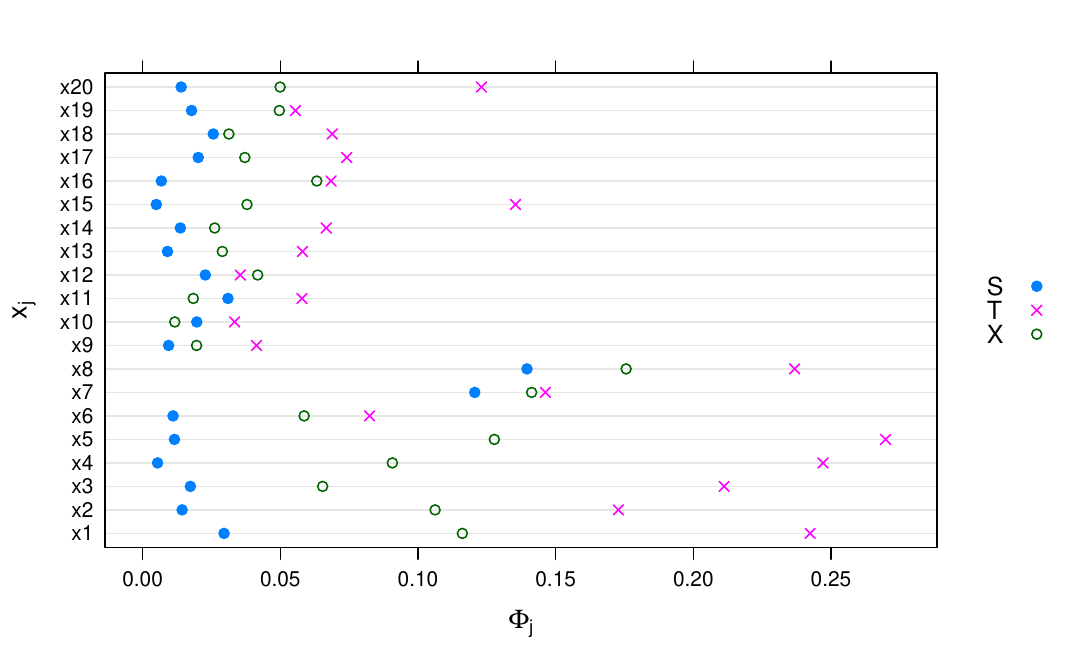}
    \caption{Illustration of SHAP rankings for three meta-learners as discussed in Section \ref{sec.head2head} where $x_7$ and $x_8$ are the predictive covariates. The main point is that SHAPs cannot meaningfully be compared across CATE models for any fixed covariate, but the inherent rankings can.}
    \label{fig.ThreeMetas}
\end{figure}

\subsection{Computational limitations of model-agnostic SHAP\label{sec.compBurden}}
The illustrative simulation example above was quite simplistic, featuring only five candidate baseline biomarkers. Despite this simplicity and a moderate sample size, the computations required to derive model-agnostic SHAP values were quite memory-intensive. During one step of the \texttt{shapr} run, when processing the initial $600 \times 5$ data, an internal data table with $656806$ rows was generated, capturing information about all possible $2^5 = 32$ coalitions of the five biomarkers. Predictions (i.e., predicting CATE) for this large dataset were necessary to compute the KernelSHAP values. This highlights a common drawback of Shapley methodology: a large number of model evaluations is often required. The authors of the \texttt{shapr} package acknowledge this in their corresponding paper\cite{Aas}, stating, ``... \emph{the kernel estimator suffers greatly from the curse of dimensionality, which quickly inhibits its use in multivariate problems}'' and ``... \emph{when the model contains more than a few covariates, computing the right-hand side of (6) is still computationally expensive}...'' While they discuss potential solutions, our testing still appears very memory-demanding. In typical randomized clinical trials, encountering $p = 20$ baseline covariates is not uncommon, which exacerbates this issue in simulation studies. With $p = 20$, there are over ten million possible coalitions $(2^{20})$ to consider, compared to just 32 with five biomarkers. Consequently, simulations quickly become infeasible for model-agnostic SHAP (KernelSHAP), especially when multiple iterations are needed to assess discovery rates. 
The same issue does not arise with the model-specific approach, because the optimization in how TreeSHAP is rendered for tree based models such as XGBoost. While the latter does require some time-consuming hyperparameter tuning it still is feasible for the settings considered in this paper, where the number of baseline biomarkers is $20$, a realistic number in many clinical trials. (High dimensional data such as omics data is not routinely  collected and/or analyzed). We have noted no memory issues when deriving SHAPs from such models.

\subsection{Comparing different CATE models using SHAPs\label{sec.head2head}}
To illustrate some further aspects of SHAP ranking assessments, data was generated from the following simulation model. The baseline covariates were 
$x_j \sim N(0,1)$ i.i.d., for $j=1,...,20$ with 1:1 treatment assignment ($P(A=1) = 0.5$), and furthermore the outcome generated as $y=-1\!+ 3\cdot (\!x_1\!+\!x_2+\!x_3+\!x_4+\!x_5)  +\!I(x_7>0, x_8>0 )\cdot\!A + \epsilon$ with  $\epsilon \sim N(0, 0.5)$. Sample size per arm was $n=400$. Of the $20$ covariates in total, only two are truly modifying the treatment effects. Therefore, an effective SHAP analysis should yield large summary SHAP values for $x_7$ and $x_8$, while all other covariates should have small SHAP values, clearly distinguishing the predictive and prognostic covariates. Such ideal results are rarely observed in finite datasets, but a correct ranking would show a clear margin between the two truly predictive covariates and the others. It is important to note that although covariates  $x_1, \ldots, x_5$ enter the simulation model, they do not contribute to any differential treatment effects and should not significantly influence CATE away from the average TE. Ideally, these covariates should receive SHAP values similar to those of non-informative covariates. If they do receive higher SHAP values, it may indicate an imperfect CATE model, suggesting mis-calibration of the prognostic effects. The results are presented in Figure \ref{fig.ThreeMetas} for the selected CATE learners T-, S- and X-learning, using XGboost as base learner (fitted by standard hyperparameter cross-validation). In the context of the strategies discussed in Section \ref{sec.strategies}, the simulations in this section follow Strategy 3.

It is important to note which aspects of SHAP values are directly comparable across different CATE estimators and which are not. By equation  (\ref{eq.AdditiveSH}) and the fact that each estimator may have its own range of values (due to inherent biases), the corresponding SHAPs mimicking model predictions will also conform to these ranges. This artifact is apparent in our example: T-learner had its largest SHAP above $0.25$ while S-learner never crossed $0.15$, i.e., their total ranges differ.  The resulting rankings of covariates can still be directly compared, while the absolute values cannot. For example, the SHAP value for $x_8$ was $\Phi^{(T)}_8=0.237$ for T-learning, while S-learner yielded $ \Phi^{(S)}_8=0.140$. If we were to consider only these values, one might be tempted to conclude that T-learner identified $x_8$ more clearly than the S-learner since $0.237>0.140$. However, a comprehensive analysis of the SHAP values for all covariates is essential to avoid such simplistic interpretations. For instance, T-learner assigned an even higher value $(\Phi^{(T)}_5=0.270)$ to a non-predictive (but prognostic) covariate $x_5$, while the S-learner assigned values $\leq 0.031$ to all other non-predictive covariates. This suggests that the ranking by T-learner is misleading in an inferential sense, implying that $x_5$ is the most important covariate. In contrast, the S-learner approach provided a more accurate ordering, ranking $x_8$ and $x_7$ above the other covariates. Similarly, X-learning correctly top-ranked the same covariates, although the separation from the other variables was less compelling in this run. 

This simulation highlights the importance of having appropriate evaluation  metrics that we will introduce in Section \ref{sec.metrics}. Beyond the interest in identifying the top-ranked biomarker, it is crucial to capture at least one truly predictive covariate when narrowing down candidates, such as the top three. It appears beneficial to employ a method that, on average, maximizes the separation between SHAP values for true positive and true negative covariates. In this simulation, T-learner exhibited a negative margin, as indicated by $\Phi^{(T)}_5 > \max(\Phi^{(T)}_7, \Phi^{(T)}_8)$. To fairly compare margins between learners (e.g., T- versus X-learner), a normalization of each set of SHAPs to the same scale is required (e.g., to $[0,1])$.  
 
\section{Large-scale simulation benchmarking}\label{sec.LSB} 
While the simulations in the previous section were included for illustrative purposes, here we perform a large-scale benchmarking of our ability to correctly identify predictive biomarkers via SHAP values. We focus on the following questions:
\begin{description}
    \item[Does the meta-learner choice affect performance of SHAP  biomarker discovery? ] Here, we aim to compare the various SHAP approaches derived from different meta-learners. Additionally, we will evaluate these approaches in both RCT and observational scenarios (Section~\ref{sec.RankingBenchmarking}).
    \item[How the prognostic strength affects the performance of the different methods? ] It has been demonstrated\cite{Hermansson2021, INFOPLUS} that some predictive biomarker discovery methods suffer more in the presence of multiple prognostic covariates (although such research did not consider SHAP). In a sense, such a presence implies a multiplicity issue, increasing the risk of incorrectly creating splits on prognostic covariates, which can lead to artifact imbalances across treatment arms within the subgroups defined by the splits. The presence of multiple prognostic biomarkers is also very realistic in clinical trial settings, where a number of prespecified prognostic covariates is usually included in data modeling (Section \ref{sec.perturb}).

    \item[How SHAP methods perform in comparison to model-specific VIP?] Here we will compare  SHAP values vs model-specific VIPs derived by Causal Forest. This is not meant to be exhaustive research on this theme, but it may shed some light on the usefulness of SHAP values (Section \ref{sec.VIPvsSHAP}).

    \item[Do the strategies for deriving SHAP for R- and DR-learner differ?] Here we will compare the two approaches suitable for the meta-learners that derive pseudo-observations: the Strategy 2 from Section \ref{sec.strategies}  deriving SHAPs when we regressing the pseudo-observations to the covariates, and Strategy 3 (surrogate strategy) deriving SHAPs by regressing the CATE estimates to the covariates (Section \ref{sec.IndirectvsDirect}). 
    \item[Can SHAP values help us to derive the marginal predictive effect?]  Apart from benchmarking the standard summary-level SHAP, we also investigate the different levels of SHAP values, available both on the performance of the instance-level importance (enabling partial dependence plots); using the same simulation model, the question becomes: which model's instance-level SHAP values recover the true marginal predictive effects? (Section \ref{sec.Finer}).
\end{description}
Next we present the simulation setup that will address these questions (Section \ref{sec.S2S3}) and the measures we will use to evaluate the performance of different analytic strategies (Section \ref{sec.metrics}).
\subsection{Simulation setup\label{sec.S2S3}}
The simulation setup used for benchmarking is similar to the one used in Lipkovich et al.\cite{ILDS}, presenting a range of challenges for the CATE approaches due to the inclusion of multiple prognostic covariates and the overall prognostic effect being stronger than the differential treatment effects. This scenario is realistic in RCT settings, where risk modeling is more routine (e.g., in designing trials to include patients more likely to have worse outcomes). 
Furthermore, the differential effects are gradual rather than sharp, and they enter in a non-monotone, non-linear fashion. Additionally, there is a subgroup of patients for whom the true treatment effect is zero.

In a similar way with Lipkovich et al.\cite{ILDS}, two  simulation models will be used, referred to as models \texttt{S2} (RCT) and \texttt{S3} (observational). These models have the same structure in the outcome model but differ in allocation to treatment: in the former model it is randomized to 3:1, while in the latter one the expected active/control balance becomes 1/3, where patients with a poor prognosis at baseline have higher likelihood to receive active. 

For the S2 model the outcome is generated by the following expression:
\begin{equation}
    y=100+k_1(\textbf{x})+k_2(\textbf{x}) \times A + \epsilon \label{eq.yK1K2}
\end{equation}
with $\epsilon \sim N(0,1),$ and $k_1(\textbf{x})$ the prognostic part given by the expression
\[
k_1(\textbf{x})=-(x_1+5x_2)+ 2(x_5+x_6+x_7+x_8+x_9),\] while the treatment-modifying part $k_2(\textbf{x})$ given by 
\begin{equation}
k_2(\textbf{x})=g_1(x_3)+g_2(x_4). \label{eq.PredPartS2S3}
\end{equation}
The covariates $X_3$ and $X_4$ are truly predictive and drive CATE in a non-linear way (and non-monotone, for $X_3$). We suppressed the index $i$ for subjects for easier readability. Here $A \sim \texttt{Bern}(0.75)$ is the binary treatment assignment resulting in a $3:1$ randomization (more patients on active). The biomarkers $X_1, X_3, X_4, ..., X_9$ are continuous $\sim N(0.5, 1)$, $X_2$ is categorical with three levels generated with equal probabilities $p=1/3$. Additional ten biomarkers are simulated as $X_j \sim N(0,1)$ for $j=10,...,19$, i.e., in total there are $19$ baseline biomarkers available as candidates for the model building.  The non-linearity in CATE is due to the transformations $g_1(\cdot)$    
\begin{equation}\label{eq.g1}
  g_1(x) = 
     \begin{cases}
      a-b\cdot 0.25 & \text{if} \;\; x < 0\\
      a-b(x-0.5)^2 & \text{if} \;\; 0 \le x \le 1 \\
      a-b\cdot 0.25 & \text{if} \;\; x > 1\\
    \end{cases},       
\end{equation}
and $g_2(\cdot)$
\begin{equation}\label{eq.g2}
  g_2(x) = 
     \begin{cases}
      0 & \text{if} \;\; x < 0\\
      \frac{c}{1+\text{exp}(-d(x-0.5))} & \text{if} \;\; 0 \le x \le 1 \\
      c & \text{if} \;\; x > 1\\
    \end{cases}.       
\end{equation}
with constants $a=0.625,b=5, c=0.625, d=20$ calibrated to render a slightly positive overall effect $E[\tau(X)]=0.0119$ where $\tau(X)$ denotes the true CATE. Assuming larger outcomes are good for the patients, the differential effects are gradual but a true meaningful subgroup of patients truly benefiting from active treatment exist: $S_{true}=\{\tau(X)>0\}$, which has prevalence $E[I(X \in S_{true})]=0.330$ and true mean effect $E[\tau(X)|X \in S_{true}]=0.665$. 

Based on the simulation model S2, we constructed scenario S3 mimicking observational data when treatmen assignments are based on  propensity scores. The model is identical to S2 except that the treatment assignment is driven by the prognostic part of the outcome model:
\begin{equation}\label{eq.ps}
 \text{logit} \left( \text{Pr}(A=1|X=\textbf{x}) \right)=\alpha_1+\beta_1 \times k_1(\textbf{x}).
\end{equation} 
where $\alpha_1=-2.4$ and $\beta_1=-0.2$ were calibrated to ensure a 1:3 ratio for treated to control subjects (which also results in sufficient overlap in the distribution of propensity across the treatment and control arms). The choice of coefficients imply the realistic scenario that patients with a poorer prognosis on the standard of care treatment have a higher chance to be assigned to the active treatment.


\subsection{Evaluation metrics\label{sec.metrics}}
In the benchmarking of operating characteristics for different CATE models, some evaluation metrics are required to assess the success of the SHAP inference. These are intended to capture to what extent truly predictive biomarkers were distinguished from truly non-predictive ones in the assessment, and this can be measured in several ways: is the top-ranked covariate truly predictive, is at least one of the top three truly predictive, and by how large margin is the separation? 

Let $X_1,..., X_p$ denote baseline biomarkers and assume there are some variable importance values associated with each biomarker: $s_j= s_j(X_j) \in \mathbb{R}$ for $j=1,...,p$. These importance measures will later often be the summary level SHAP importance associated to the biomarker $X_j$: 
\begin{equation}
s_j = \frac{\sum_{i=1}^{n} |\phi^i_{j}|}{n}, \label{eq.defSumSHAP}
\end{equation} as defined in Section \ref{sec.SHAP_THEORY} but will occasionally be other types (model specific) of variable importance (e.g., VIP from Causal Forest). Given the vector $(s_1,...,s_p)$, a ranking of the biomarkers is inherent and given via ordering the $s_j$, which is denoted by
\[s^{(o)}= \big(s^{(1)}, s^{(2)},..., s^{(p)}\big)\] where $|s^{(j)}| \geq |s^{(k)}|$ whenever $1 \leq j<k \leq p$. Let furthermore $(X^{(1)}, X^{(2)},..., X^{(p)})$ denote the corresponding order of the biomarkers associated with ordered vector $s^{(o)}$. To examplify with $p=3$ covariates, assume $s_1=0.11$, $s_2=7.29$ and $s_3=2.02$; this renders $s^{(o)}=(7.29,  2.02, 0.11)$. In this example, $X^{(1)}=X_2, X^{(2)}=X_3$ and $X^{(3)}=X_1$ and the resulting importance ranking would be $X_2 \gg X_3 \gg X_1$, where $a \gg b$ means ``$a$ is  ranked higher than $b$''. To clarify, the ranking disregards the numeric importance values and instead focuses on the order of priority for further investigations into potential differential treatment effects; for example, \emph{`study $X_2$ first'} in the toy scenario.

Therefore, given the above, there is a mapping  
\[ M^{(o)}: (X_1, ..., X_p) \rightarrow (X^{(1)}, X^{(2)},..., X^{(p)}) =(X_{i_1}, X_{i_2},...,X_{i_p}),
\] or, equivalently, a mapping from original indices to new ones: $M^{(o)}: (1,...,p) \rightarrow (i_1, i_2,...,i_p)$. (In the toy example above, $i_1=2, i_2=3, i_3=1$). Now, to introduce the metrics, denote  \texttt{P} the set of truly predictive biomarkers and \texttt{NP} 
 the complementary set of all non-predictive ones (i.e., merely prognostic and/or non-informative ones; any covariate being both prognostic and predictive is counted as `predictive'). To exemplify, \texttt{P}=$\{3, 4\}$  and 
 \texttt{NP}=$\{1, 2, 4, ..., 19\}$ in the simulation model S2 in Section \ref{sec.S2S3}. Then the following metric tracks status of the top-ranked covariate: 
\[
\texttt{TOP}_{1}^{\texttt{(M)}}= 
\begin{cases}
    1,  & \text{if } max\{|s_j|; j\in \texttt{P}\} \geq max\{|s_j|; j\in \texttt{NP}\}\\
    0,              & \text{otherwise}
\end{cases}
\]
Furthermore, denote $L^{(3)}=\{i_1, i_2, i_3\}$ the set of indices for the covariates with three highest importance measures; then, 
\[
\texttt{NET}_{3}^{\texttt{(M)}}= 
\begin{cases}
    1,  & \text{if }  \texttt{P}\cap L^{(3)} \neq \emptyset \\
    0,              & \text{otherwise}
\end{cases}
\] which is $1$ whenever a truly predictive covariate is captured in the top-3 ranking.
The third metric is defined as 
\[\texttt{MARGIN}^{\texttt{(M)}}= 
    max\{|s_j|; j\in \texttt{P}\} - max\{|s_j|; j\in \texttt{NP}\}
\] which measures the amount of separation between predictive and non-predictive covariates inherent in the importance measures.  As a final note, different CATE models typically render different overall range and distribution of estimates, resulting in SHAPs having different overall ranges, as follows from equation (\ref{eq.AdditiveSH}). (This was illustrated in Section \ref{sec.head2head})). Therefore, for each CATE model, we first standardize the SHAPs  to the range $[0,1]$ prior to any comparison of \texttt{MARGIN}.


\subsection{Does the meta-learner choice affect performance of SHAP  biomarker discovery?\label{sec.RankingBenchmarking}}
\begin{figure}
    \centering
    \includegraphics[scale=0.8]{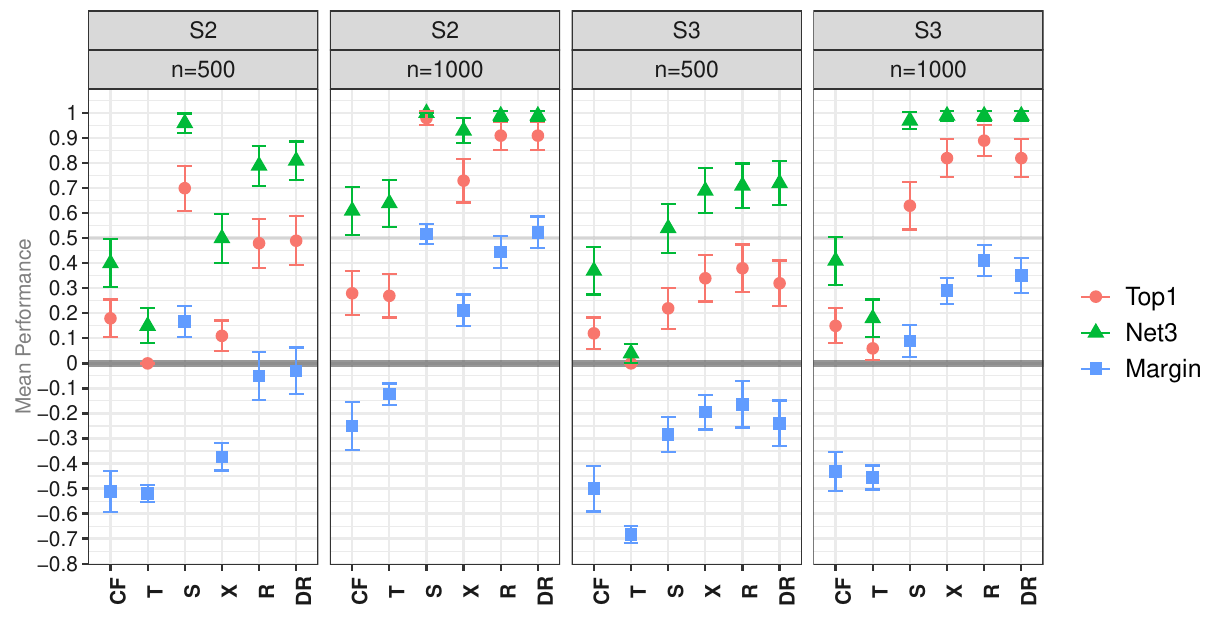}
    \caption{Individual results (over 100 simulation runs) from large scale benchmarking of the impact of meta-learner choice on SHAP biomarker discovery performance, evaluated across three metrics.} 
    
    \label{fig.MetricsPlot}
\end{figure}

In this section, we present results regarding our ability to distinguish between predictive and non-predictive biomarkers using the surrogate SHAP (Strategy 3). To explore this question we generated data/results in the following way. Training data is simulated from scenarios S2 and S3 (sample size $n=500, 1000$). Each type of CATE-model (CF, T-, S-, X-, R-, DR-learner) is fitted to the data. For each such model, surrogate SHAPs are derived for each of the biomarkers. The \texttt{TOP1}, \texttt{NET3}, and \texttt{MARGIN} metrics (see section \ref{sec.metrics} for details) are derived from the SHAPs for each CATE model, recording how well predictive biomarkers were identified. We repeated this process for a $100$ iterations, the average results are listed in Table \ref{tab.MetricTab} and individual runs are displayed as box plots in Figure \ref{fig.MetricsPlot}.

Our first insight is that in the S2 simulation model (RCT), the S-learner performs the best across all evaluation measures and sample sizes, with a runner-up being the DR-learner. It is worth mentioning that the T-learner has a very poor performance. The conclusions differ for the S3 simulation model (observational data), where the meta-learners that model the propensity score, such as the R-learner, DR-learner, and X-learner, outperform the other methods. Overall, the Causal Forest has poor performance across all scenarios.

\begin{table}[ht]
\caption{Average result from large scale benchmarking of the ability to recover predictive biomarkers, in three metrics. \label{tab.MetricTab}}
\centering
\begin{tabular}{cccccc}
  \hline
Simulation Model & n & CATE estimator & Top1 $\pm$  SE & Net3 $\pm$ SE & Margin $\pm$ SE \\ 
  \hline
\multirow{6}{*}{S2} & \multirow{6}{*}{500}  & T & 0.00 $\pm$ 0.000 & 0.15 $\pm$ 0.036 & -0.52 
$\pm$ 0.018 \\ 
&  & S & \textbf{0.70} $\pm$ 0.046 & \textbf{0.96} $\pm$ 0.020 & \textbf{0.17} $\pm$ 0.031 \\ 
&  & X & 0.11 $\pm$ 0.031 & 0.50 $\pm$ 0.050 & -0.37 $\pm$ 0.028 \\ 
&  & CF & 0.18 $\pm$ 0.038 & 0.40 $\pm$ 0.049 & -0.51 $\pm$ 0.042 \\ 
&  & R & 0.48 $\pm$ 0.050 & 0.79 $\pm$ 0.041 & -0.05 $\pm$ 0.049 \\ 
&  & DR & 0.49 $\pm$ 0.050 & 0.81 $\pm$ 0.039 & -0.03 $\pm$ 0.048 \\ \hdashline 
\multirow{5}{*}{S2} & \multirow{6}{*}{1000}  & T & 0.27 $\pm$ 0.044 & 0.64 $\pm$ 0.048 & -0.12 $\pm$ 0.022 \\ 
&  & S & \textbf{0.98} $\pm$ 0.014 & \textbf{1.00} $\pm$ 0.000 & \textbf{0.52} $\pm$ 0.020 \\ 
&  & X & 0.73 $\pm$ 0.044 & 0.93 $\pm$ 0.026 & 0.21 $\pm$ 0.032 \\ 
&  & CF & 0.28 $\pm$ 0.045 & 0.61 $\pm$ 0.049 & -0.25 $\pm$ 0.050 \\ 
&  & R & 0.91 $\pm$ 0.029 & 0.99 $\pm$ 0.010 & 0.44 $\pm$ 0.032 \\ 
&  & DR & 0.91 $\pm$ 0.029 & 0.99 $\pm$ 0.010 & \textbf{0.52} $\pm$ 0.032 \\ \hdashline 
\multirow{5}{*}{S3} & \multirow{6}{*}{500}  & T & 0.00 $\pm$ 0.000 & 0.04 $\pm$ 0.020 & -0.68 $\pm$ 0.017 \\ 
 &  & S & 0.22 $\pm$ 0.041 & 0.54 $\pm$ 0.050 & -0.28 $\pm$ 0.036 \\ 
&  & X & 0.34 $\pm$ 0.047 & 0.69 $\pm$ 0.046 & -0.19 $\pm$ 0.036 \\ 
&  & CF & 0.12 $\pm$ 0.032 & 0.37 $\pm$ 0.048 & -0.50 $\pm$ 0.047 \\ 
&  & R & \textbf{0.38} $\pm$ 0.049 & 0.71 $\pm$ 0.045 & \textbf{-0.16} $\pm$ 0.048 \\ 
&  & DR & 0.32 $\pm$ 0.047 & \textbf{0.72} $\pm$ 0.045 & -0.24 $\pm$ 0.046 \\ \hdashline 
\multirow{5}{*}{S3}& \multirow{6}{*}{1000}  & T & 0.06 $\pm$ 0.024 & 0.18 $\pm$ 0.038 & -0.46 $\pm$ 0.024 \\ 
&  & S & 0.63 $\pm$ 0.048 & 0.97 $\pm$ 0.017 & 0.09 $\pm$ 0.033 \\ 
&  & X & 0.82 $\pm$ 0.038 & \textbf{0.99} $\pm$ 0.010 & 0.29 $\pm$ 0.027 \\ 
&  & CF & 0.15 $\pm$ 0.036 & 0.41 $\pm$ 0.049 & -0.43 $\pm$ 0.040 \\ 
&  & R & \textbf{0.89} $\pm$ 0.031 & \textbf{0.99} $\pm$ 0.010 & \textbf{0.41} $\pm$ 0.031 \\ 
&  & DR & 0.82 $\pm$ 0.038 & \textbf{0.99} $\pm$ 0.010 & 0.35 $\pm$ 0.036 \\ 
   \hline
\end{tabular}
\end{table}

\begin{figure}
    \centering
    \includegraphics[scale=0.8]{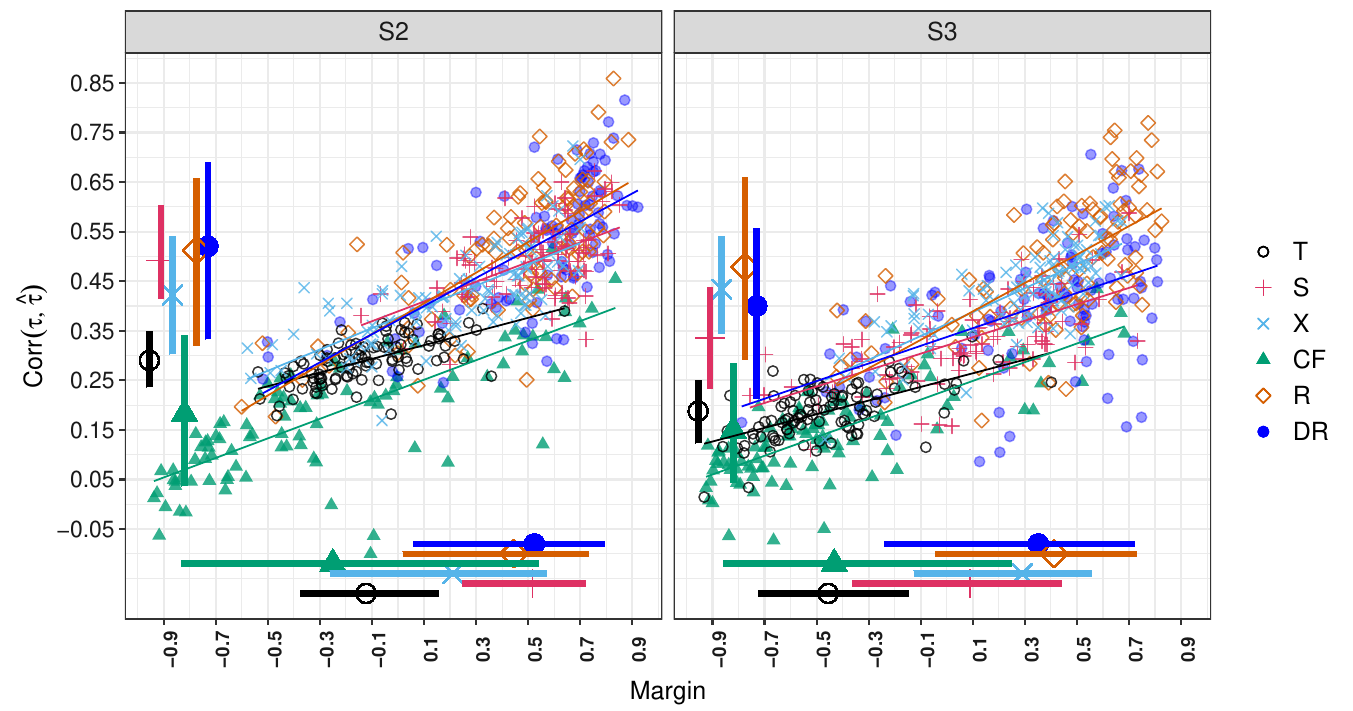}
    \caption{Large scale benchmarking of the relationship between the ability to distinguish predictive and prognostic biomarkers (measured by \texttt{MARGIN}) and the quality of CATE estimation. The latter is assessed by the correlation between true and estimated CATE values ($cor(\tau, \hat{\tau})$), plotted on the y-axis. An exploratory linear regression was fitted to the predictions  for each model, highlighting the relationship between CATE estimation quality and biomarker ranking accuracy. 
    The simulations were conducted using models S2 and S3 with a sample size of $1000$. The bars show 10th and 90th percentiles.}
    \label{fig.NEWCORRSHAPmargin}
\end{figure}

One interesting question is how the ability to correctly estimate CATE is correlated with the ability to distinguish between predictive and prognostic biomarkers. To explore this question we generated data/results in the following way. Training data is simulated from scenarios S2 and S3 (sample size $n$). Each type of CATE-model is fitted (CF, T-, S-, X-, R-, DR-learner) to the data. For each such model, surrogate SHAPs are derived for each of the biomarkers. The  \texttt{MARGIN} metric is derived from the SHAPs for each CATE model. An independent test dataset is simulated (sample size $10000$), and each model is used to predict CATE on test data resulting in $\hat{\tau}$ for each modeling choice. The correlation of $\hat{\tau}$ and the simulation truth $\tau$ is then recorded for each method, measuring how well CATE was estimated in the sense of correctly ranking patients regarding their true individual treatment effect. We repeated this process for a $100$ iterations, the results are presented in Figure \ref{fig.NEWCORRSHAPmargin}. 

Along the x-axis is the \texttt{MARGIN} metric, measuring the amount of separation between predictive and non-predictive covariates. For example, in the left panel (S2) the right-top-most blue filled circle (DR-learner) shows one iteration where the \texttt{MARGIN}  was approximately 0.9 (x-axis) and the correlation was around 0.80. A positive \texttt{MARGIN} indicates that at least one of the truly predictive covariates $x_j$ received a higher summary SHAP value than all non-predictive $x_j$'s (hence top-ranked) and the larger \texttt{MARGIN}  is, the better the separation; i.e., the clearer the identification of one truly predictive biomarker. Therefore, in this figure it is advantageous to be in is the upper right corner.  The solid lines are globally fitted linear regression showing the overall trend.

Figure~\ref{fig.NEWCORRSHAPmargin} demonstrates a strong positive correlation between the ability to rank biomarkers correctly (measured by \texttt{MARGIN}) and the quality of CATE estimation (measured by the correlation between true and estimated CATE values, $cor(\tau, \hat{\tau})$). The exploratory linear regression analysis further supports this relationship, showing a clear trend, which indicates that models providing more reliable CATE estimates also tend to have higher accuracy in ranking biomarkers. More specifically, in the left panel (model S2), we observe that the DR-learner method exhibits the highest performance on both axes, underscoring its effectiveness in both biomarker ranking and CATE estimation. In contrast, the right panel (model S3) shows that the R-learner method achieves the highest performance. These findings underscore the importance of accurate biomarker ranking in enhancing the quality of CATE estimation.


\subsection{How does the prognostic strength affect the performance of the different methods? \label{sec.perturb}}
To answer this question, we make a slight modification to the model setup presented in Section \ref{sec.S2S3}: we introduce an overall prognostic strength factor $\beta \geq 0$, turning the outcome model into 
\begin{equation}
y=100+ \beta\cdot k_1(\textbf{x})+k_2(\textbf{x})\cdot A + \epsilon. \label{eq.newK1K2}
\end{equation}
For the observational version of this model (S3), the assignment to active is kept as before (i.e., ensuring the 1:3 balance with more controls). 
The choice $\beta = 1$ obviously corresponds to the original model given by expression (\ref{eq.yK1K2}), while $0 \leq \beta < 1$ implies relatively stronger differential treatment effects, and likewise  $\beta > 1$ implies relatively stronger prognostic effects. The latter case is likely to make the detection of $x_3$ and $x_4$ harder, and a research question is if it matters what CATE model is used to this end? The same iterative setup will be used as in simulation section \ref{sec.RankingBenchmarking}, but with the small extra detail that in any single iteration, the same baseline data and residual error is simulated once only and held fixed for generating the outcome $Y$ for different candidate choices for $\beta \in [0,  0.5, 1,  1.5, 2]$. The number of iterations is $100$, and this time the sample size is fixed to $n=1000$. Again, all the CATE models considered before again are applied, surrogate SHAPs generated (Strategy 3), and metrics tracked and summarized. 

\begin{figure}
    \centering
\includegraphics[scale=0.7]{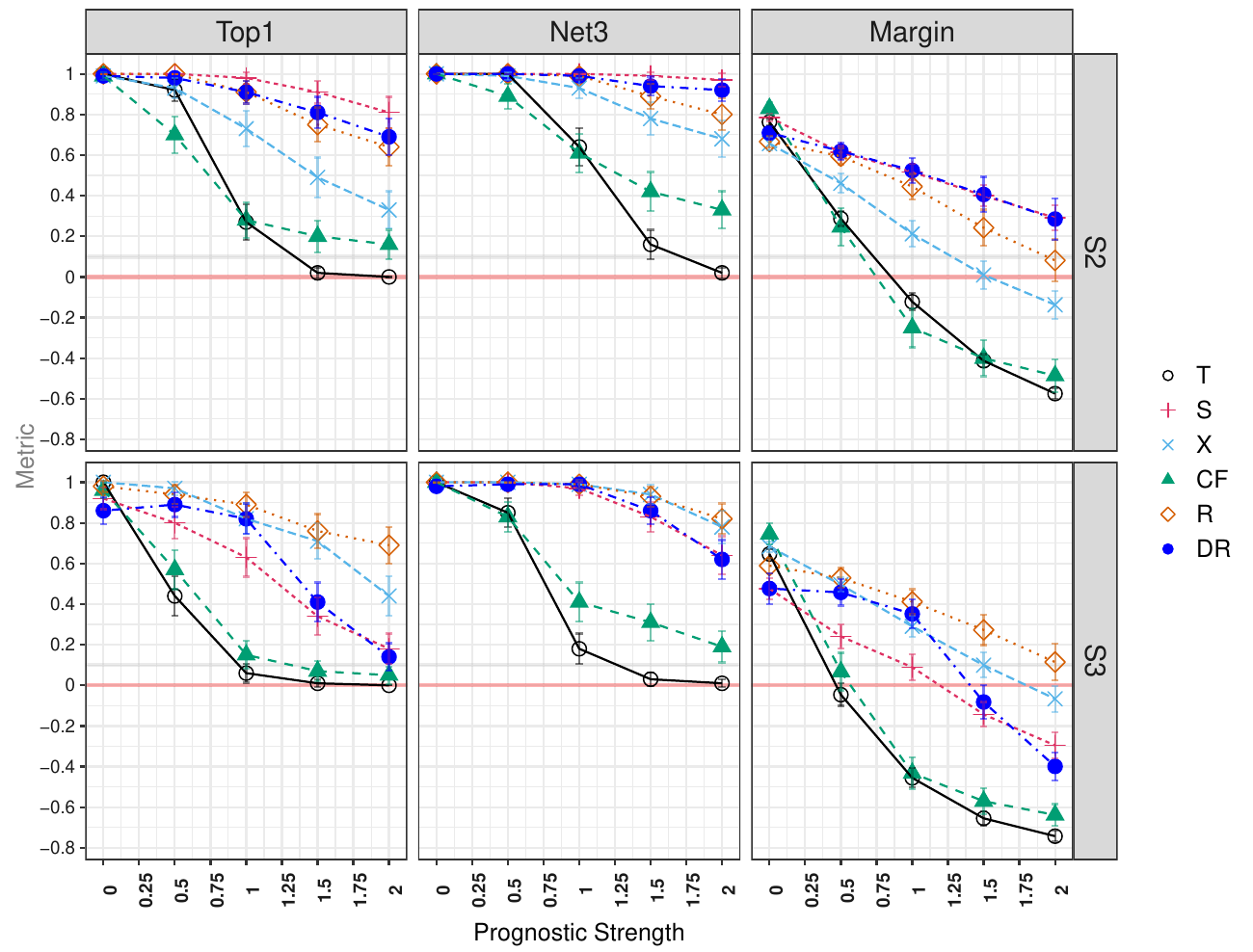}
\caption{Averaged performance ($100$  iterations, sample size $n=1000$) plotted against overall true prognostic strength $\beta$ in the outcome model $y=100+ \beta\cdot k_1(\textbf{x})+k_2(\textbf{x})\cdot A + \epsilon$. See Section \ref{sec.perturb}) for details. The graph shows how well competing CATE models are able to distinguish between predictive and prognostic covariates, as the strength of the latter increase. The horizonal grey line at $y=0.10$ represents performance of   random guessing (regarding the \textbf{$TOP_1$} metric), and notably the T-learner and Causal Forest drop below this threshold for some configurations.}
    \label{fig.perturbFig}
\end{figure}

The results are displayed in Figure \ref{fig.perturbFig} where it is notable that the popular T-learner deteriorates faster than the other approaches. This finding is in line with other research\cite{Hermansson2021} and indicates that arm-specific modeling is prone to misunderstanding prognostic effects since the arm-specific models don't communicate (and hence can arrive at disagreement regarding main effects, since these are estimated twice despite being constant across treatment arms). Interestingly, T-learner is numerically better-or-equal than the other approaches when there is a complete absence of prognostic effects ($\beta = 0$). It is worth noting that the \texttt{NET3} metric does not improve much over \texttt{TOP1} for T-learning; of the five truly prognostic covariates, suggesting that that they are all top-ranked, hence dominating the top-3 SHAP selection. The performance of T-learner in terms of \texttt{MARGIN} was off the charts and negative already at $\beta= 0.5$, meaning there is less than $50$\% chance to capture a truly predictive biomarker in the SHAP top-ranking. 

The other approaches perform better in this sense due to either using common modeling across arms (Causal Forest, S-learner), or other regularization techniques (X-, R- DR-learner). It is assuring that the modern approaches X-, R- and DR-learner land on relatively high probability to net at least one truly predictive biomarker even when the prognostic strength is very high (middle panel in Figure \ref{fig.perturbFig}, $\beta =2$).  For the RCT setting (model S2) the S- and DR-learner perform the best, while for the observational setting (model S3) the X- and R-learner outperform the rest of the approaches. 

Finally, it is worth noting that performing worse than randomly guessing is possible. Random guessing strategy will by chance correctly top-rank either $x_3$ or $x_4$ (both truly predictive) $10\%$ of the time (counting here dummy coding for $x_2$ as two separate covariates, making it $20$ candidate covariates in total), but our empirical results suggest worse results are possible: in the left-most panel in Figure \ref{fig.perturbFig} three approaches land on \texttt{TOP1} $< 0.10$ (for $\beta =2$) while the T-learner arrives there already at $\beta = 1$. This may look counter-intuitive but is caused by the ``increasing focus'' on prognostic covariates as their strength grows.

\subsection{How SHAP methods perform in comparison to model-specific VIP? \label{sec.VIPvsSHAP}}
In this section we illustrate that SHAP based ranking of biomarkers may improve over standard method-specific variable importance (VIP). Many different versions of VIP for supervised learning is  established in the data science area, e.g. Gini Index and permutation-based VIPs for RandomForest\cite{RandomForest} and Relative Influence for Gradient Boosting\cite{XGB_paper}. The research here is not meant to be exhaustive regarding the landscape of possible VIP implementations, but rather to illustrate that there might be good reasons to consider SHAP over default model-specific choices also in the setting of estimating individual treatment effects. Here, we therefore focus on Causal Forest (CF) with its built in VIP metric which are directly targeting the predictive rather than prognostic effects\cite{ILDS_ClinTrials}. The CF VIP measure is a weighted sum of the number of times each covariate was split on at each depth in the forest\cite{GRF}. The simulation model was the S2 model with a total sample size of $n=1000$. The performance is averaged over $500$ iterations, each consisting of generating data, fitting CF (including full tuning via \texttt{tune.parameters = ``all"} argument) and rendering the CF-VIP and surrogate SHAPs, i.e. Strategy 3 combined with CF (more details in Section \ref{sec.strategies}). 

Only two of the biomarkers are  truly predictive in our setup, and hence an ideal ranking should have these as top 1-2.  Metrics \texttt{TOP1} and \texttt{NET3} were applied both to SHAP and CF-VIP.   Averaged result across iterations are found in Table \ref{tab.CFvipvsshap}, clearly suggesting that SHAPs are more accurate and stable. E.g., there is a considerable $57.5\%$ improvement in the probability that the top-ranked covariate actually is truly predictive.  

\begin{table}[ht]
\centering
\caption{Comparison of model-specific VIP and SHAP methods using CATE estimation from Causal Forest, averaged over 500 iterations. \label{tab.CFvipvsshap}}
\begin{tabular}{cccc}
  \hline
Importance Type &  TOP1 $\pm$  SE(TOP1) & NET3 $\pm$ SE(NET3) & MARGIN $\pm$ SE(MARGIN) \\ 
  \hline
  SHAP & \textbf{0.316} $\pm$ 0.0208 & \textbf{0.550} $\pm$ 0.0222 & \textbf{-0.015} $\pm$ 0.0033 \\ 
VIP & 0.196 $\pm$ 0.0178 & 0.398 $\pm$ 0.0219 & -0.045 $\pm$ 0.0029 \\ 
   \hline
\end{tabular}
 \end{table}

\subsection{Do the strategies for deriving SHAP for R- and DR-learner differ? \label{sec.IndirectvsDirect}}
The surrogate SHAP approach (Strategy 3) was introduced to offer a unified way of deriving SHAP regardless of how CATE was estimated, but specific implementations may offer other possibilities. In this section we compare this approach with Strategy 2, which is suitable for reducible methods such as the R- and DR-learner (more details in Section \ref{sec.strategies}). These meta-learners follow a two-step strategy, where in the first step a pseudo-observation is derived ($\widehat{\psi}_{R}$ and $\widehat{\psi}_{DR}$ respectively) and then a final model is build to estimate CATE.  

\begin{figure}
    \centering
\includegraphics[scale=0.9]{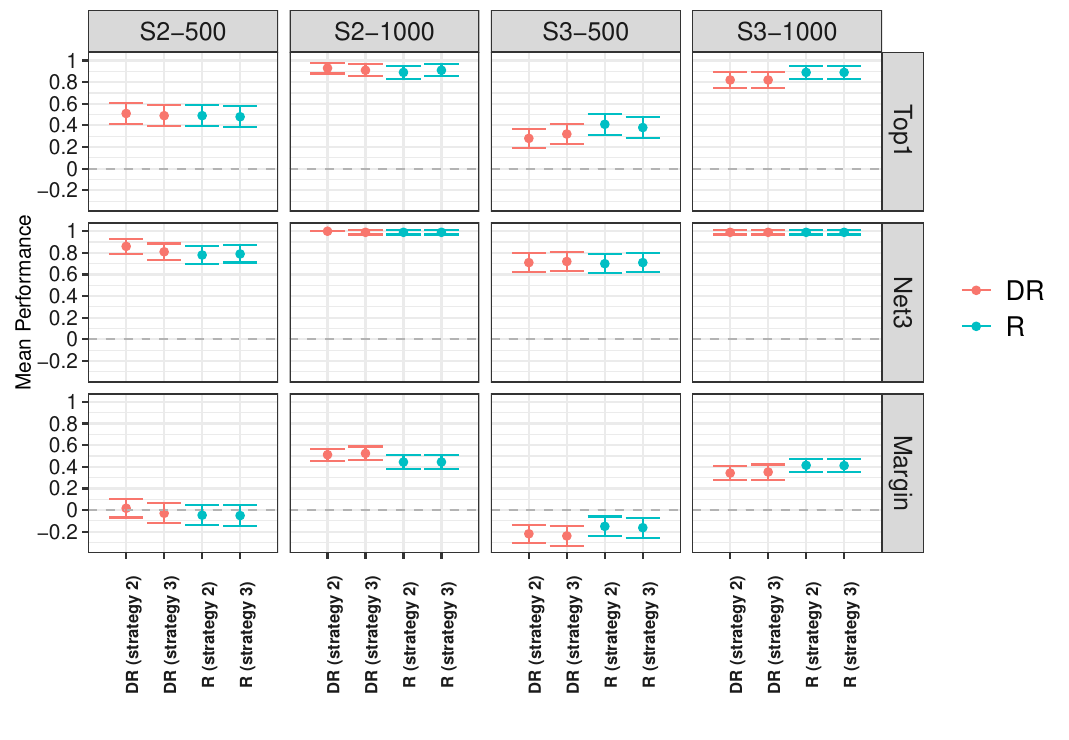}

    \caption{Comparison of averaged performance across three metrics (100 iterations) for two strategies of deriving SHAP values in R- and DR-learners: Strategy 2 (relying on their inherent reducibility as discussed in Section \ref{sec.strategies}, deriving SHAP by regressing the pseudo-observations, i.e.  $\hat{\psi} \sim \textbf{x}$) and the surrogate method (deriving SHAP from the model that regresses on the CATE estimates, i.e.  $\hat{\tau} \sim \textbf{x}$).}
    \label{fig.CheckDR_R_2SHAPs}
\end{figure}

A natural question is to consider if these more direct approaches are superior in performance to the surrogate approach, given the avoidance of an extra modeling step? To assess this, data was simulated in $100$ iterations from models S2 and S3 (sample size $n=1000$) as  described in Section \ref{sec.perturb}. Results are summarized in 
Figure \ref{fig.CheckDR_R_2SHAPs} which presents averaged results in the three metrics for models S2 and S3, and for sample sizes 500 and 1000. 
No empirical evidence was found to indicate a statistical difference in performance between the two types of SHAPs. The surrogate approach offers pragmatic simplicity, making it applicable across various CATE models without sacrificing performance.

\subsection{Can SHAP values help us to derive the marginal predictive effect? \label{sec.Finer}}
In this section we look beyond the summary SHAPs and consider some inference relying on the finer granularity of SHAPs. As described in section \ref{sec.SHAP_THEORY}, the Shapely/SHAP idea is suitable not only to rank covariates on their overall importance, but also to ``explain'' individual decisions taken by the predictive model. In CATE estimation, this issue goes beyond just ranking covariates; it involves the analysis of instance-level SHAPs to see if they can provide insights into systematic variations in CATE. This is clearly a central question in precision medicine: whenever there is some evidence for treatment effect heterogeneity in drug development (e.g, supported by some combination of analytic assessment of the current data, domain knowledge, biological plausibility, and maybe external data) it becomes greatly important to also infer potential relationships, looking for trends and potential cut points on biomarkers (which further supports precision medicine designs and decisions). It is therefore motivated to explore the potential of using instance-level SHAP values for this purpose. However, other model-dependent local measures, such as exploring Partial Dependence Plots for XGBoost \cite{xgboostpackage}, could also be considered.

\begin{figure}
    \centering
\includegraphics[scale=0.7]{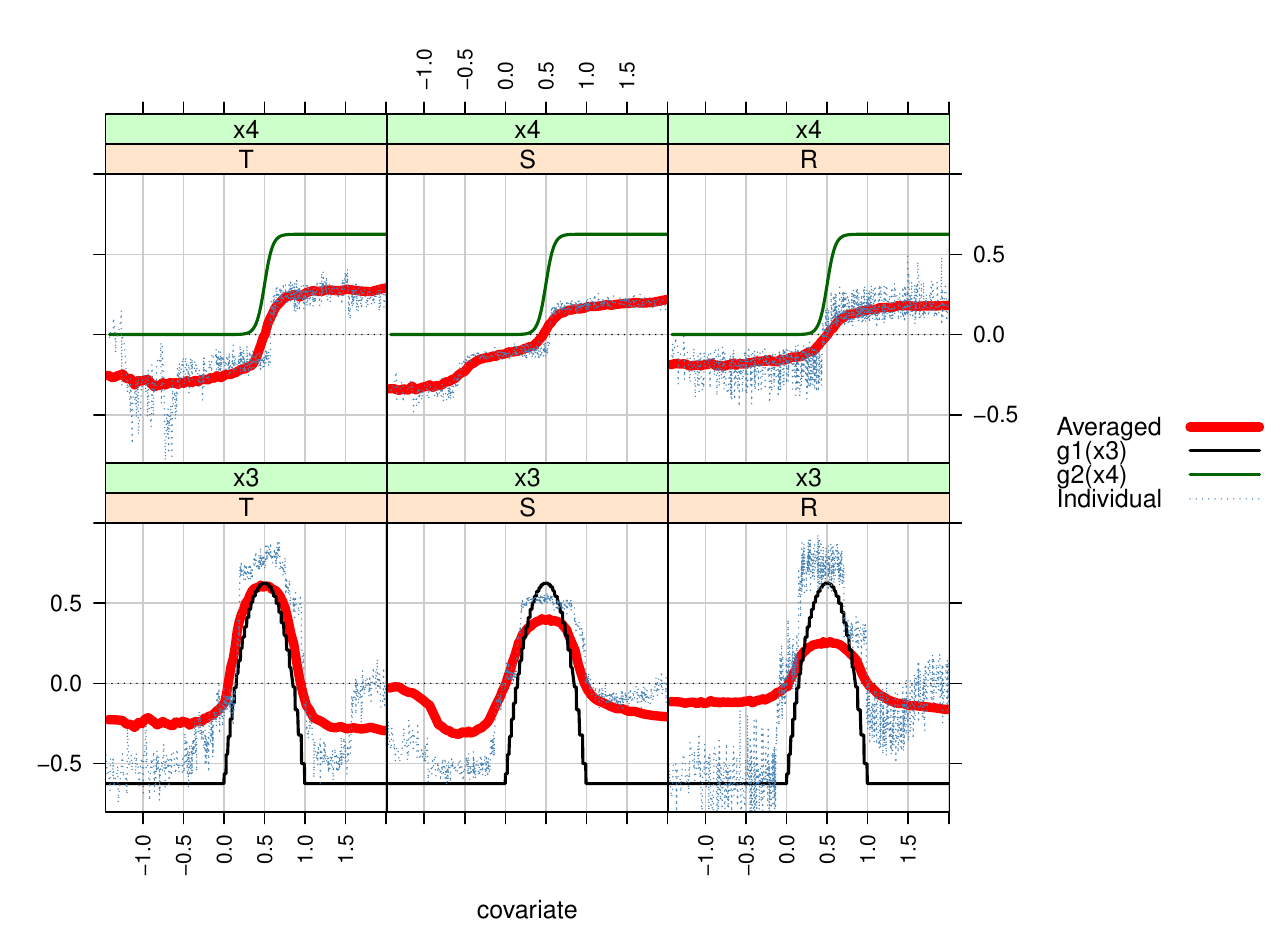}
\caption{Illustration of recovery of $g_1(x_3)$ and  $g_2(x_4)$ using local SHAPs in simulation model S2 ($n_{tot}=1000$). Results from selected CATE models are displayed (T-, S- and R-learning): blue dotted line shows SHAP from a single simulation iteration; red line is the averaged result over $100$ iterations. 
}
    \label{fig.Recovery1}
\end{figure}

We will explore these questions using our main simulation model (Section \ref{sec.S2S3}). First, the ability to recover the true functional shapes $g_1(\cdot)$ and $g_2(\cdot)$ using local SHAP values will be considered. This is intended as a proof-of-concept demonstration, assuming there is a priori knowledge about which covariates are truly predictive, i.e., not taking into account the other covariates. Instance-level SHAP profiles for the truly predictive covariates are displayed with $g_1(x_3)$ and $g_2(x_4)$ in Figure \ref{fig.Recovery1}.  The profiles do not include the SHAP intercept $\phi^i_0$ from equation \ref{eq.AdditiveSH} motivated by the following.  Imagining a perfect CATE model, it would archive $\hat{\tau}(\textbf{x}_i)= \tau(\textbf{x}_i)=\{$by definition in Section \ref{sec.S2S3}$\}= g_1(x_{1,i})+g_2(x_{2,i})$, and have SHAP for non-predictive covariates equal to zero, i.e., $\phi^i_k=0$ for $k \neq \{3,4\}$. Therefore, $g_1(x_{1,i})+g_2(x_{2,i})= \phi^i_0 + \phi^i_3 + \phi^i_4$. In the following, we chose to ignore the overall shift $\phi^i_0$ and focus on how well local SHAPs  qualitative can recover the simulation truth. 

The oracle truth is overlaid in the graph (green line for $g_2(x_4)$ and black line for $g_1(x_3)$) In this graph we plot (blue dotted lines) SHAP values $\hat{\phi}_3$ and $\hat{\phi}_4,$ for the covariates $x_3$ and $x_4,$ in the lower and upper panel respectively. Although the SHAP values for only one individual iteration are shown in this figure, it demonstrates the inherent promise of SHAPs and how there is a clear tendency to qualitatively capture the underlying relationships.  While the blue line represents the data that the operator would typically have in a practical situation, in our simulation setup we have the luxury to observe the average trend across 100 iterations (red line) as well. The figure demonstrates a clear potential to capture the underlying true signal, as both the red and blue lines reflect the cut-points and trends, despite not aligning perfectly with the black line. Additionally, it is important to note the significant differences in performance among the various CATE models.

To explore even further the ``quality'' of these instance-level SHAPs we averaged Pearson correlations $|cor(\hat{\phi}_3, g_1(x_{i,3}))|$  and $|cor(\hat{\phi}_4, g_2(x_{i,4}))|$ for the different CATE learners across $100$ iterations, with results summarized in Table \ref{tab.CORRSHAPgFUN}. Causal Forest, is performing considerably worse than the meta-learners, e.g., correlations regarding $x_3$ were $0.357$ and $0.368$ for S2 and S3, respectively, far below the other approaches. T-learner performed strongly, in particular for scenario S2. It is however instructive to contrast this with what happens when a selection problem is present, as now follows. 

As a background to the following investigation,  the true correlations between the $g()$-functions and true CATE are high: $cor(\tau, g_1(x_{3}))=0.843$, $cor(\tau, g_2(x_{4}))=0.537$ in our simulation model (as empirically derived on data with $n= 10^6$), while all other correlations are zeros, i.e. $cor(\tau, x_k)=0$ where   $k \ne \{3,4\}$. 
To consider how often false discovery can occur, we track how often the highest $cor(\hat{\phi}_j, \widehat{\tau}(x))$ appears for a truly non-predictive covariate (i.e., $j \neq \{3,4\}$), i.e., occurring by chance only. Some additional notation is needed to formalize some metrics to compare the different methods on their quality of instance-level SHAP values. 

Let us denote with $\texttt{P}=\{3,4 \}$ the set of indices of predictive covariates and \texttt{NP} = the set associated with the non-predictive ones. Also, let $\hat{\phi}_j=\hat{\phi}^i_j$, $i=1, \dots, n$ be the instance-level SHAPs assigned to covariate $j$ from the surrogate approach, and let $\hat{\rho}_j = |cor(\hat{\phi}_j, \widehat{\tau}(x)|$, the Pearson correlation. In each iteration, the following is tracked: $max(P)=max \{\hat{\rho}_j; j=\texttt{P} \}$ and $max(NP)=max \{ \hat{\rho}_j; j=\texttt{NP}\}$. In practice, inspection of instance-level SHAP profiles is likely to focus on covariates with high $\hat{\rho}_j$, hence reflecting a selection of a biomarker.

Similar to earlier metrics, we introduce an instance-level margin defined as $max(P) - max(NP)$, where a negative value indicates a false discovery. In our simulation, we average the success indicator function $I = I_{max(\texttt{P}) \geq max(\texttt{NP})}$, denoted as $P_{win}$ in the output (see Table \ref{tab.recoveryTab}). This tracks how prone different CATE approaches are to discover the wrong covariate. In this setup, the S-learner performs the best in the $S2$ scenario, with the DR-learner as the runner-up. Conversely, in the $S3$ scenario, the R-learner performs the best, with the DR-learner again as the runner-up. Interestingly, the T-learner performs poorly, despite showing strong results when considering only $x_3$ and $x_4$ (Table \ref{tab.CORRSHAPgFUN}). This indicates its tendency to select prognostic covariates as predictive, leading to spurious local SHAP patterns, as evident in our results. 

To summarize, the above illustrations highlight the inherent difficulties in inspecting individual SHAP profiles in practical problems. ``Interesting patterns'' may easily appear for the wrong reasons. Our empirical investigation suggests considering one of the better-performing meta-learners for such ``data insights,'' such as the R-learner or DR-learner.

\begin{table}[ht]
\centering
\caption{Results regarding ability to recover the oracle truth $g_1(x_3)$ and $g_2(x_4)$ for the simulation model S2 ($n=1000$) over $100$ iterations, see Section \ref{sec.Finer} for details. (Notation: cor3=$|cor(\hat{s}_{i,3}, g_1(x_{i,3}))|$  and cor4=$|cor(\hat{s}_{i,4}, g_2(x_{i,4}))|$, standard errors in parentheses)\label{tab.CORRSHAPgFUN}}
\begin{tabular}{cccc}
  \hline
scenario & method & cor3 $\pm$ SE(cor3) & cor4 $\pm$  SE(cor4) \\ 
  \hline
\multirow{6}{*}{S2} & T & \textbf{0.867} $\pm$  0.006 & 0.815 $\pm$  0.013 \\ 
 & S & 0.850 $\pm$  0.006 & 0.829 $\pm$  0.012 \\ 
 & X & 0.830 $\pm$  0.008 & \textbf{0.861} $\pm$ 0.014 \\ 
 & CF & 0.357 $\pm$  0.024 & 0.597 $\pm$  0.052 \\ 
 & R & 0.768 $\pm$  0.013 & 0.820 $\pm$  0.020 \\ 
 & DR & 0.761 $\pm$  0.015 & 0.851 $\pm$  0.022 \\ \hdashline
\multirow{6}{*}{S3} & T & 0.722 $\pm$  0.014 & 0.688 $\pm$  0.017 \\ 
 & S & 0.729 $\pm$ 0.013 & 0.792 $\pm$  0.012 \\ 
 & X & \textbf{0.832} $\pm$  0.012 & \textbf{0.861} $\pm$  0.010 \\ 
 & CF & 0.368 $\pm$  0.020 & 0.596 $\pm$  0.045 \\ 
 & R & 0.746 $\pm$  0.016 & 0.809 $\pm$  0.013 \\ 
 & DR & 0.648 $\pm$  0.021 & 0.799 $\pm$  0.021 \\ 
   \hline
\end{tabular}

\end{table}

\begin{table}[ht]
\centering
\caption{Average results regarding inference using the instance-level SHAPs, here assessing the strategy to selecting the covariate with highest SHAP-correlation with CATE. Simulation model was $100$ iterations, and P(win) is the empirical average across iterations regarding $I_{max(\texttt{P}) \geq  max(\texttt{NP})}$. See Section \ref{sec.Finer} for details regarding the definition of P and NP, and consider \ref{sec.perturb}) for the simulation models S2 and S3; n=1000the table is for prognostic strength case $\beta=1$ and sample size $n=1000$. \label{tab.recoveryTab}}  
\begin{tabular}{cccccc}
  \hline
 scenario & n & method & max(P) & max(NP) & $P_{win}$ $\pm$  SE($P_{win})$ \\ 
  \hline
\multirow{6}{*}{S2} & \multirow{6}{*}{1000} & T & 0.36 & 0.445 & 0.03 $\pm$  0.017 \\ 
 &  & S & 0.53 & 0.370 & \textbf{0.97} $\pm$  0.017 \\ 
 &  & X & 0.46 & 0.413 & 0.64 $\pm$  0.048 \\ 
 &  & CF & 0.38 & 0.514 & 0.26 $\pm$  0.044 \\ 
 &  & R & 0.62 & 0.434 & 0.85 $\pm$  0.036 \\ 
 &  & DR & 0.54 & 0.253 & 0.92 $\pm$  0.027 \\ \hdashline
\multirow{6}{*}{S3} & \multirow{6}{*}{1000} & T & 0.33 & 0.471 & 0.01 $\pm$  0.010 \\ 
 &  & S & 0.42 & 0.415 & 0.52 $\pm$  0.050 \\ 
 &  & X & 0.48 & 0.415 & 0.76 $\pm$  0.043 \\ 
 &  & CF & 0.33 & 0.559 & 0.14 $\pm$  0.035 \\ 
 &  & R & 0.62 & 0.441 & \textbf{0.81} $\pm$  0.039 \\ 
 &  & DR & 0.46 & 0.292 & 0.80 $\pm$  0.040 \\ 
   \hline
\end{tabular}
\end{table}


\section{Case studies}
\label{sec.CASE}
\subsection{Case study 1: Application to a clinical trial on schizophrenia }\label{sec.CASE1}
In this section, we revisit the case study from a recent paper\cite{ILDS}, which involves a synthetic dataset based on a Phase III clinical study for a treatment of schizophrenia. Originally a 4-arm RCT, only the highest dose ($n=149$) and placebo ($n=152$) groups are considered for illustration. The outcome variable is the change from baseline in the Positive and Negative Syndrome Score at Day 42 (PANSS-42) total score, a common endpoint in schizophrenia studies, with negative values indicating health improvement. The dataset contains eight baseline covariates: two categorical (\texttt{Gender}, \texttt{Race}), one ordinal (\texttt{cgis}), and five continuous (\texttt{age}, \texttt{diagyears}, \texttt{pansspos}, \texttt{panssneg}, \texttt{panssgen}).

As described in the aforementioned paper, this data represents a common setting where the overall effect is estimated to be close to zero (p-value = $0.615$), prompting potential analyses to investigate if some subpopulations might still benefit from the active drug. In this particular dataset, it is unlikely\cite{ILDS} that any substantial heterogeneity of treatment effect is present. It is instructive to consider what SHAP values from different CATE models reveal here. The approaches from Section \ref{sec.CATE} were applied to the data, and surrogate SHAP values were derived. The methods slightly disagreed regarding the top-ranked covariate but ultimately pointed to either \texttt{pansspos} or \texttt{diagyears}, as shown in Table \ref{tab.topRankedSHAPSchizo}.
\begin{table}[ht]
\centering
\caption{Case Study 1: Top-ranked covariate across CATE models, according to global SHAPs.\label{tab.topRankedSHAPSchizo}}
\begin{tabular}{rllr}
  \hline
 what & Top1 & shap.max \\ 
  \hline
 C & pansspos & 0.51 \\ 
   T & diagyears & 2.01 \\ 
   S & diagyears & 0.89 \\ 
  X & diagyears & 0.92 \\ 
   R & pansspos & 0.11 \\ 
   DR & pansspos & 0.61 \\ 
   \hline
\end{tabular} 
\end{table}
\begin{figure}[htb] 
\centering
\begin{subfigure}[b]{0.40\textwidth}
    \centering
 \includegraphics[width=7cm, height=6.5cm]{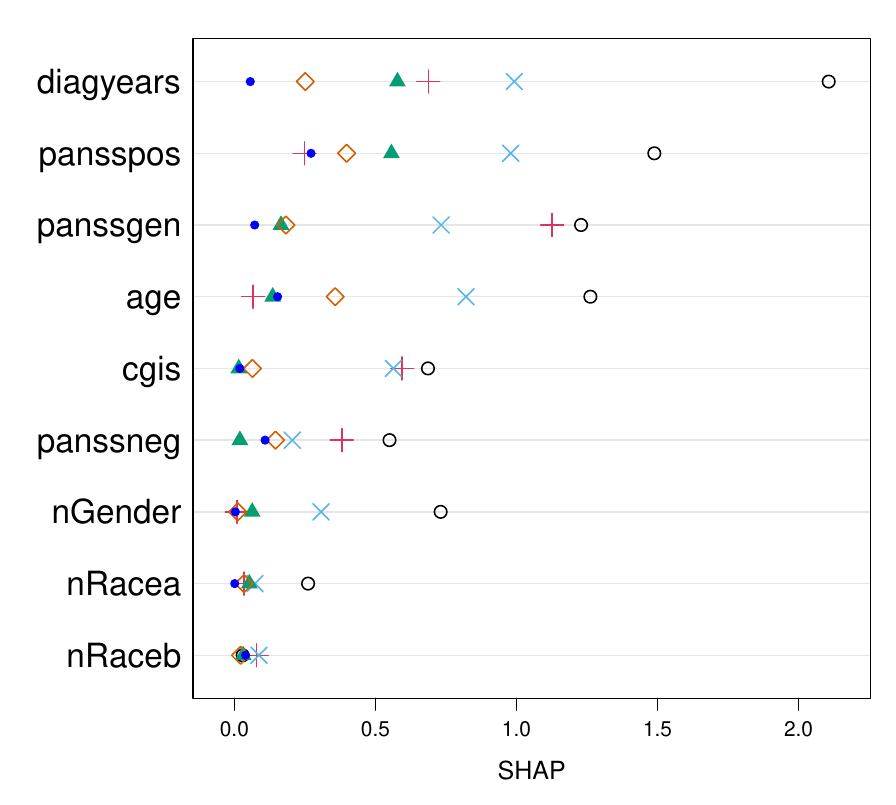}
\end{subfigure}
\begin{subfigure}[b]{0.50\textwidth}
    \centering
 \includegraphics[width=8cm, height=6.5cm]{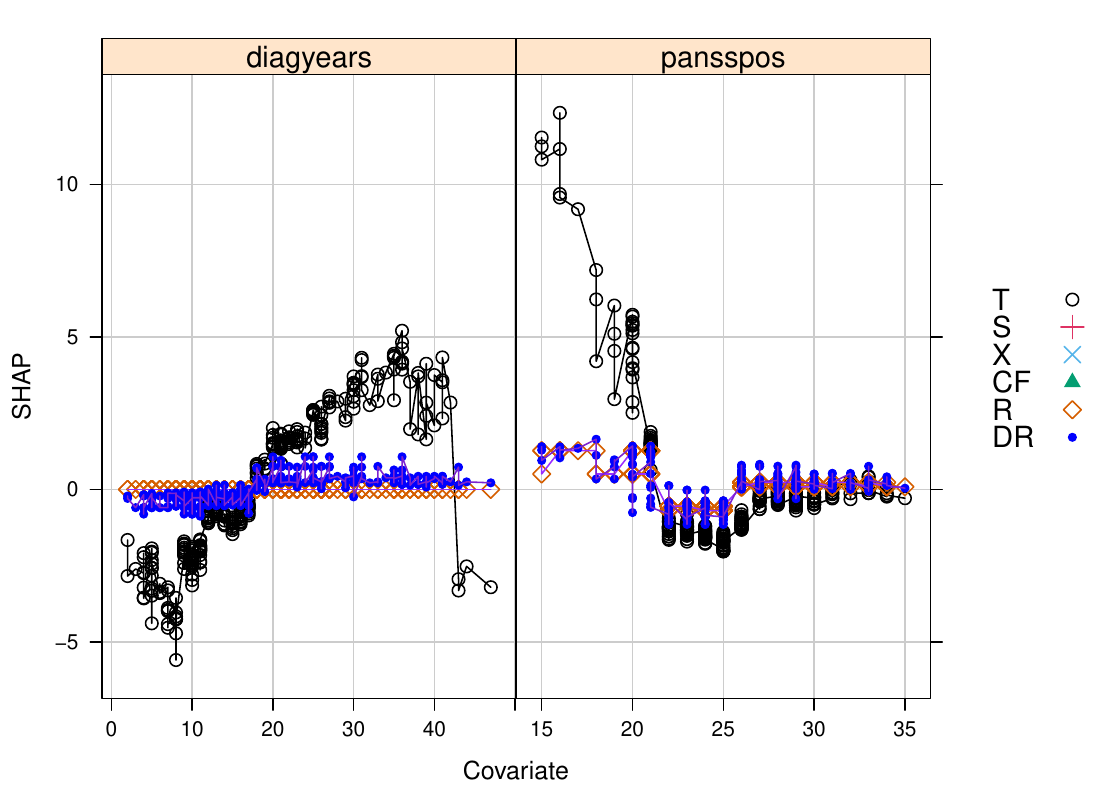}
\end{subfigure}
\caption{Case study 1: the schizophrenia trial. Left panel shows the overall ranking (y-axis). The Local SHAP profiles for selected CATE models are displayed for the two top-ranked covariates. Other approaches not shown to maintain clarity in the visualization. \label{fig:RCT_CASE1}
}
\end{figure}

This closely replicates the ranking from earlier investigations\cite{ILDS}, where VIPs from Causal Forest preferred \texttt{Age} over \texttt{Pansspos}. Their application of the \texttt{SIDESscreen} method replicated our top-ranking result, as judged by an aggregation over the different CATE models (see Figure \ref{fig:RCT_CASE1}, left panel): \texttt{Diagyears} followed by \texttt{Pansspos}. To further illustrate attempts to inspect local SHAP for trends and cut-points, the results from T-, R-, and DR-learning are overlaid in the right panel. Here, T-learning produces striking patterns, seemingly suggesting the presence of meaningful cut-points on the covariates. Other methods disagree, displaying less pronounced patterns for R- and DR-learning. Based on the performance results from our benchmarking study, this casts some doubt on the T-learning results.

\subsection{Case study 2: Application to acupuncture headache trial}\label{sec.CASE2}
For illustrative purposes, we revisit a clinical trial data example from a recent methodological paper on Random Forest Interaction Trees\cite{Su}. The actual data is publicly available for download at \url{https://trialsjournal.biomedcentral.com/articles/10.1186/1745-6215-7-15}. This randomized clinical trial on chronic headache (predominantly migraine) involved 401 patients randomized to either receive acupuncture treatments for three months or usual care. The sponsor of the trial acknowledged its limitations (lack of blinding\cite{ACU}), but it still provides a practical example of applying CATE-learners, which can be contrasted with the example in\cite{Su}. The primary endpoint was the change from baseline in headache score (SF-$36$) at 12 months.
\begin{figure}[htb] 
 \begin{subfigure}[b]{0.4\textwidth}
    \centering
    \includegraphics[scale=0.55]{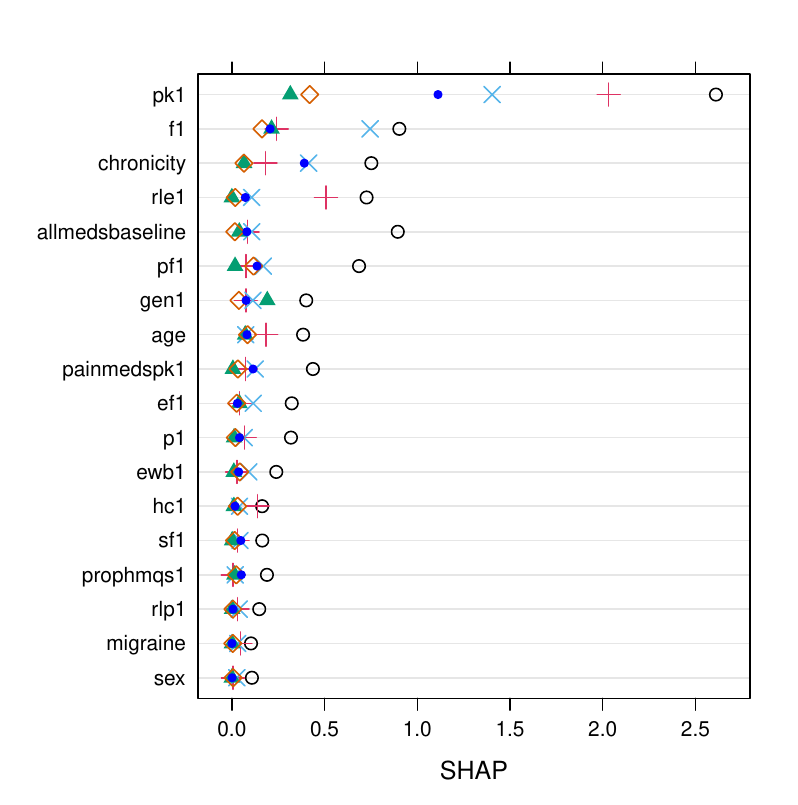}    
\end{subfigure}
 \begin{subfigure}[b]{0.5\textwidth}
    \centering
    \includegraphics[scale=0.55]{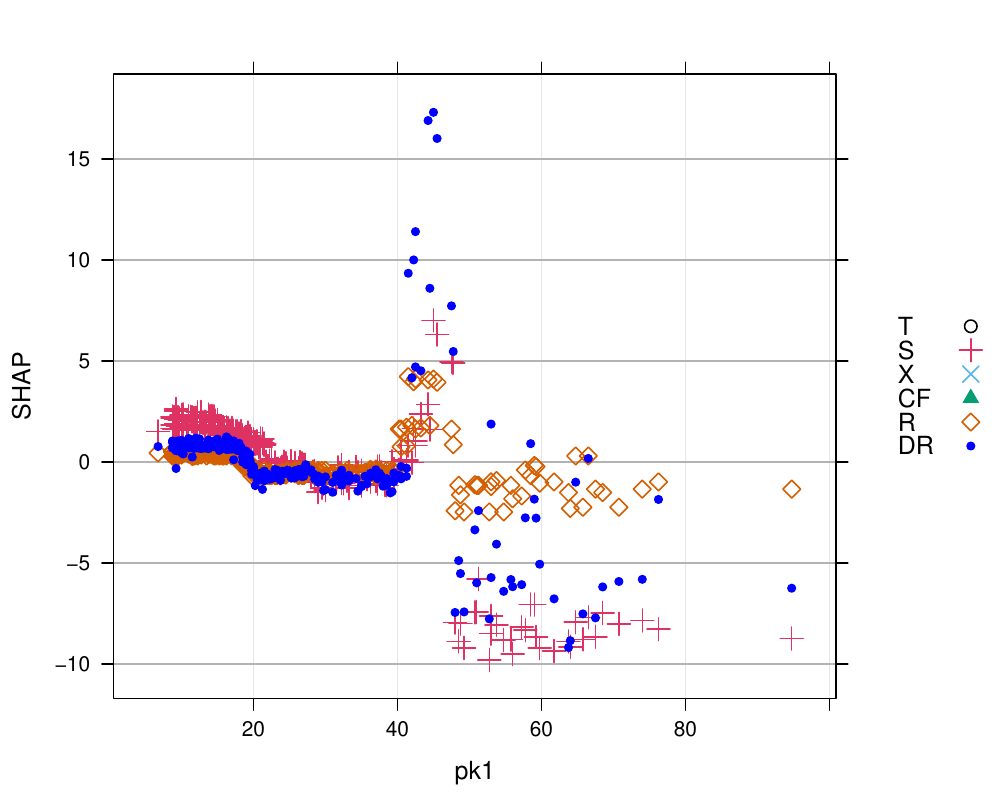}
\label{fig:CASE2_global}
 \end{subfigure}
    \caption{Case Study 2: the acupuncture headache trial. Left panel shows the overall ranking (y-axis). The Local SHAP profiles for selected CATE models are  displayed for the top-ranked \texttt{Pk1} covariate in the right subplot (S-, R- and DR-learning). Similar results was qualitatively obtained for the other approaches, although the magnitudes differed across methods (e.g., all SHAPs for Causal Forest were all close to zero, as also visible in the summary ranking plot).  The profiles suggest a meaningful cut point around $45$.}\label{fig:CASE2_local}
\end{figure}

We followed \cite{Su} in the data preparation, analyzing $301$ participants who completed the trial (i.e., drop-out effects were ignored here). The endpoint is the difference in headache severity score between baseline and 12 months, with the baseline severity score included as one of the $18$ baseline covariates available in this study. These covariates were: demographics (\texttt{age}, \texttt{sex}); disease status (\texttt{migraine}, \texttt{f1}, \texttt{chronicity}); disease severity score (\texttt{pk1}); multi-item scales for SF36 (\texttt{pf1}, \texttt{rlp1}, \texttt{rle1}, \texttt{ef1}, \texttt{ewb1}, \texttt{sf1}, \texttt{p1}, \texttt{gen1}, \texttt{hc1}); Medication Quantification Scale (\texttt{painmedspk1}); MQS of prophylactic medication (\texttt{prophmqs1}); and Total MQS (\texttt{allmedsbaseline}). The resulting assignment to treatments was $n=140$ for the control arm and $n=161$ for the active arm. The overall effect, adjusted for baseline, replicates the reported\cite{Vickers} $4.586$ units improvement in disease severity score, favoring active treatment.

According to Su et al.\cite{Su}, there was only moderately strong evidence for treatment effect heterogeneity. The magnitude of the averaged ITE (mean of estimated CATE) was reported as $3.9$ (or $-3.9$ in the placebo-adjusted sense), matching our mean estimates (e.g., DR-learning renders -2.25, and -3.18 for S-learning). While they focused on individual estimates (e.g., reporting that $76.85\%$ of the patients were predicted to benefit from the treatment), our focus is on the identification of predictive covariates. SHAP values for the top-ranked baseline covariate are displayed in Figure \ref{fig:CASE2_local}, left panel, for the six CATE models. All these approaches generally agree on the overall ranking, with the covariate `pk1' (severity score at baseline) being top-ranked. The SHAP patterns suggest that the reduction in severity score depends sharply on the baseline score. Selecting a meaningful cut-point to define a subgroup for another trial is a topic in itself and outside the scope of this work, but likely candidate splits might be found around $45$ (depending on additional criteria regarding effect size, uncertainty, and prevalence).

To exemplify what a more classical subgroup discovery approach might yield, we ran MOB\cite{Seibold2016}, starting with only treatment as a covariate. This resulted in two splits on $\texttt{pk}_1$, the first at $28.5$, with higher values on $\texttt{pk}_1$ associated with higher treatment effect estimates. While this split point differs slightly from what is visually suggested in the SHAP profile plot (right panel in Figure \ref{fig:CASE2_local}), it qualitatively agrees that this covariate might interact with treatment. Another ad hoc exploratory result further supports this impression: the interaction p-value is $p=8.705e-05$ from the model $\texttt{Pk5}i=S_i + A_i + S_i\cdot A_i + \epsilon_i$, with $S_i=I(\texttt{pk1}{i}>45)$ being a subgrouping factor regarding the (admittedly somewhat arbitrary) candidate split value of $45$. The $A_i$ denotes treatment assignment. This was included merely to illustrate that SHAP can provide insights into cut-points.

\section{Discussion }\label{sec.DISC} 
SHAP has quickly become a go-to solution for explaining complex machine learning models in the supervised learning setting. However, little attention has been given to SHAP values for causal models, despite the practical importance of gaining insights into their predictions. While methods for estimating conditional average treatment effects (CATE) have received considerable attention in recent years, most existing research has focused on the finer aspects of estimation, such as regularization and different ways of performing statistical tests of heterogeneity based on CATE. 
This trend has led to many different modeling schemes, often involving fitting multiple models. Some of these schemes, which involve fitting a single final model (e.g., X-learner, R-learner), allow us to directly derive SHAP values. However, establishing a unifying approach for efficient SHAP value estimation, applicable to any type of CATE estimator has remained largely underexplored.

In this tutorial we provide a road map and systematic evaluation (via benchmarking simulations) for various approaches to deriving SHAP values in the context of HTE estimation. As a unifying approach we propose a straightforward strategy that entails fitting an XGBoost model to  CATE estimates  obtained by any method against all potential effect-modifiers as covariates, and deriving SHAP values from this model. This approach mimics other established methods in the subgroup discovery literature, such as Virtual Twins\cite{Foster2011} and recent work by Kennedy\cite{Kennedy2025}. 
Our approach, termed Surrogate SHAP, replaces the simple CART models used in the aforementioned works with XGBoost, the industry-standard tree ensemble model in supervised learning and derives Tree-SHAP values from it. This is in line with the proposal in a recent book\cite{BOOKBIOMARKER}, which mentions the possibility of replacing CART with a tree ensemble to access classical variable importance (VIP). The Surrogate approach is, to our knowledge, novel and bypasses inherent questions about where exactly in a multi-stage CATE scheme to derive the SHAP values. By relying on the tree-based implementation of SHAP (TreeSHAP), we obtain exact Shapely Values and avoid the computational challenges of the (inexact) model-agnostic KernelSHAP, which is known to scale poorly with the number of covariates \cite{lundberg2020local}. 

Having established that the Surrogate approach is a worthy contender for deriving SHAP values for arbitrary CATE models, we can now explore the notion of explaining such models: what covariates contribute to the prediction of individual treatment effects, and how do they do so? In standard statistical language, this involves identifying and understanding predictive biomarkers, a key aspect of precision medicine\cite{INFOPLUS}. However, a potential difficulty arises. While SHAP's solid theoretical grounding makes it a good choice for providing insights into complex models, we should be mindful of its limitations. An important one is that although it can inform us of the contribution of each covariate to the model, the model itself can be an arbitrarily bad estimator of CATE -in which case, explaining key contributors to a poor prediction is not particularly useful. 

Evaluating a CATE model is inherently challenging due to the fundamental problem in causal inference: the target (ITE) is not directly observable. This contrasts with supervised learning, where standard resampling techniques can estimate prediction accuracy via held-out samples (e.g., cross-validation). Research on the evaluation of CATE (e.g., \cite{ImaiLi2021}) is still young, with no clear consensus on how to confidently determine which model is superior for a given dataset. This dilemma has motivated our further investigations in this paper. While SHAP has theoretically good properties, not knowing if the model was well-fitted leads to the open question of how correct the inference is. Whenever there is a striking pattern, how likely is it to be meaningful or an artifact? For example, are the top-ranked covariates likely to be truly predictive, or is there a tendency to incorrectly identify merely prognostic ones (as reported in \cite{Hermansson2021} for standard VIP measures)? Our benchmarking considered the extent to which the inherent model architecture can impact SHAP inference. Empirically, the results demonstrate a clear relationship between top-ranking quality and estimation accuracy.

The investigations in this paper are a direct follow-up to those conducted by Lipkovich et al.\cite{ILDS}, where various CATE models competed across several simulation scenarios. The focus there was on estimation properties rather than biomarker selection. Two of these scenarios were the basis for further simulations here: one RCT setup with many prognostic covariates, and one RWE scenario assuming that patients with a poor prognosis are more likely to receive active treatment. Diverse results were confirmed again, highlighting that the modeling scheme significantly impacts outcomes, which motivates ongoing research in this area. One notable finding is that two popular approaches, T-learning and Causal Forest, struggled to distinguish between prognostic and predictive covariates. T-learning erroneously assigned high SHAP values to prognostic covariates. While T-learning is valid for estimating CATE, two separate arm-specific models cannot, by design, arrive at identical conclusions regarding main effects. This issue is exacerbated when there are many effects, as in our setup. Eventually, one prognostic covariate is claimed differently by the two models, presenting it as a treatment effect modifier. In real-world data, we cannot control the strength of prognostic factors, but it was instructive to see that increasing this strength in our simulations, could make SHAP rankings worse than random guessing for T-learning. Surprisingly, though less drastically, the same issue occurred with Causal Forest, warranting further research to pinpoint the cause of this failure. We note, as stated in \cite{ILDS}, that the simulation setup was intended to be challenging, with multiple potential issues. Nonetheless, other CATE learners provided more reassuring results. The S-learner performed well in the RCT setup but was somewhat inferior to more modern approaches (X-, R-, and DR-learners) in observational data. Overall, R-learning produced the best results in our benchmarking.

These findings relate to the overall rating in terms of global SHAP values. We also explored the promise of local explanations, although benchmarking such usage of SHAP is not straightforward. We believe its strength lies in complementing other statistical biomarker analyses, offering a `second view' for possible trends and cut-points for top-ranked covariates, potentially supported by other techniques. Our simulation setup illustrated this usage, showing that local SHAP values aligned well with true relationships, albeit mostly qualitatively (modulo the SHAP intercept). Cut-points were clearly visible in our simulation examples. Interestingly, T-learning, despite its drawback of two non-communicating models, seemed the most flexible in recovering the true functions. However, this was not the whole story, as SHAP values for prognostic covariates also tended to display trends and patterns in individual iterations. We tracked how strongly local SHAP values correlated with CATE for each covariate and found that more modern CATE models were less prone to pointing to incorrect variables. S- and R-learning more often prioritized truly predictive covariates over non-predictive ones.

Although SHAP is currently a very popular model explanation paradigm in data science, many other techniques for Explainable Machine Learning have been developed, and how such alternatives would perform in head-to-head comparisons has been out of scope in our work. Another interesting direction for future work would be to go beyond the basic uses of the SHAP framework in CATE estimation. As already  mentioned, one limitation of SHAP, is that it does not account for the model's performance. An interesting alternative would be to use a loss-based (rather than a prediction-based) definition of SHAP values, i.e. rather than measuring the contribution of each covariate to the model's output (prediction) we could instead measure their contribution to the model's loss (quality of the prediction). To our knowledge a limited number of studies \cite{lundberg2020local, madakkatel2024llpowershap} have attempted this and it has not been explored in the context of CATE estimation -largely due to the aforementioned difficulty in evaluating its performance. Going beyond single-covariate effects, future work can include SHAP interaction plots for further understanding what covariates might interact with one another and with the treatment itself. Taking this a step further, SHAP analysis can be extended to account for SHAP interaction effects involving multiple variables \cite{muschalik2024shapiq}. Last but not least, a closer analysis of SHAP values across individual instances could facilitate finding meaningful cut-points on biomarkers (e.g. see the case of covariate $x_3$ in Figure \ref{fig.ModelAgnosticVSModelSpecific}, for a characteristic example). Local shap values across all covariates, can potentially also be used as a basis for identifying subgroups with enhanced effect. 




\section*{Acknowledgments}
We thank the EFSPI Treatment Effect Heterogeneity Special Interest working group (SIG) for many valuable discussions. Special thanks to Kristian Brock for proof-reading this manuscript.

\appendix

In our testing, the most general approach (Strategy 1) suffered from computational costs (memory, run times) and came with some additional noise: see Section \ref{sec.ExDirIndirT} for a worked example.  Here, we briefly discuss the usage of the kernels instead of TreeSHAP for Setting 2. Data was generated from the simulation model described  in Section \ref{sec.ExDirIndirT}, but here with additional noisy covariates added to make the total number of covariates equal to $p=100$. To estimate CATE, T-learning based on RandomForest was used, and an XGboost model with default settings and $nround=200$ trees was used as $\texttt{M}^{(2)}$, from which both tree- and kernel-based SHAP was derived. The former took $0.481$ seconds to run on a linux platform (with 32 GB RAM memory, R 4.0.2, x86$\_$64-pc-linux-gnu).
\begin{table}[ht]
\caption{Computational cost for the R package \texttt{kernelshap} in the toy example for different sampling sizes.\label{tab.cost}} 
\centering
\begin{tabular}{cc}
  \hline
 Sampling Size & Elapsed Time \\ 
  \hline
 2  & 25.41 \\ 
 10  & 54.63 \\ 
200  & 586.84 \\ 
   \hline
\end{tabular}
\end{table}
The computational times for the latter are listed in Table \ref{tab.cost}, and was magnitudes larger depending on the sampling strategy used. With up to eight covariates, the package computes exact KernelSHAP, but sampling based approximations\cite{Aas} sets in when the number of covariates is higher. In total, $100$ iterations of the above were conducted, tracking global SHAPs for both approaches; the discrimination between the predictive $x_3$ and the prognostic covariates was also worse for the kernel-based approach compared to treeSHAP.

\bibliography{references.bib} 


\end{document}